\documentclass[prd,nofootinbib,showpacs,longbibliography]{revtex4-1}
\usepackage{amssymb}
\usepackage{amsmath}
\usepackage{amsfonts}
\usepackage{hyperref}
\usepackage[noabbrev]{cleveref}

\newcommand{\sh}{{\mathcal{H}}}
\newcommand{\gc}{G_c}
\newcommand{\pmvec}{{(\pm1)}}
\newcommand{\Mp}{M_\mathrm{Pl}}
\allowdisplaybreaks

\begin{document}

\title{Inflationary Instabilities of Einstein-Aether Cosmology}
\author{Adam R. Solomon}
\author{John D. Barrow}
\affiliation{DAMTP, Centre for Mathematical Sciences, University of Cambridge, Wilberforce
Rd., Cambridge CB3 0WA, UK}
\begin{abstract}
We examine the consequences of Lorentz violation during slow-roll inflation. We consider a canonical scalar inflaton coupled, through its potential, to the divergence of a fixed-norm timelike vector field, or ``aether." The vector is described by Einstein-aether theory, a vector-tensor model of gravitational Lorentz violation. We derive and analyze the cosmological perturbation equations for the metric, inflaton, and aether. If the scale of Lorentz violation is sufficiently small compared to the Planck mass, and the strength of the scalar-aether coupling is suitably large, then the spin-0 and spin-1 perturbations grow exponentially and spoil the inflationary background. The effects of such a coupling on the CMB are too small to be visible to current or near-future CMB experiments; unusually, no isocurvature modes are produced at first order in a perturbative expansion around the aether norm. These results are discussed for both a general potential and a worked example, $m^2\phi^2$ inflation with a quadratic scalar-aether coupling term.
\end{abstract}

\maketitle
\tableofcontents

\section{Introduction}

Lorentz invariance is a cornerstone of modern physics. Two separately successful theories have been constructed upon it: general relativity (GR) to explain the structure of spacetime and gravity, and the standard model of particle physics to describe particles and non-gravitational forces in the language of quantum field theory (QFT). Each apparently contains Lorentz symmetry as a crucial underlying tenet.

What do we gain from exploring this fundamental symmetry's breakdown? Given its foundational significance, the consequences of violating Lorentz invariance deserve to be fully explored and tested. Indeed, while experimental bounds strongly constrain possible Lorentz-violating extensions of the standard model \cite{Mattingly:2005re}, Lorentz violation confined to other areas of physics --- such as the gravitational, dark, or inflationary sectors --- is somewhat less constrained, provided that its effects are not communicated to the matter sector in a way that would violate the standard-model experimental bounds. Moreover, it is known that general relativity and the standard model should break down around the Planck scale and be replaced by a new, quantum theory of gravity. If Lorentz symmetry proves not to be fundamental at such high energies --- for instance, because spacetime itself is discretised at very small scales --- this may communicate Lorentz-violating effects to gravity at lower energies, which could potentially be testable. The study of possible consequences of its violation, and the extent to which they can be seen at energies probed by experiment and observation, may therefore help us to constrain theories with such behavior at extremely high energies.

A pertinent recent example is Ho\v{r}ava-Lifschitz gravity, a potential UV completion of general relativity which achieves its remarkable results by explicitly treating space and time differently at higher energies \cite{Horava:2009uw}. The consistent non-projectable extension \cite{Blas:2009qj,Blas:2009ck,Blas:2010hb} of Ho\v{r}ava-Lifschitz gravity is closely related to the model we will explore. Moreover, since we will be dealing with Lorentz violation in the gravitational sector, through a vector-tensor theory of gravity, the usual motivations from modified gravity apply to this kind of Lorentz violation. Indeed, there are interesting models of cosmic acceleration, based on the low-energy limit of Ho\v{r}ava-Lifschitz gravity, in which the effective cosmological constant is technically natural \cite{Blas:2011en,Audren:2013dwa}. Generalized Lorentz-violating vector-tensor models have also been considered as candidates for both dark matter and dark energy \cite{Zlosnik:2006zu,Zuntz:2010jp}.

Lorentz violation need not have such dramatic, high-energy origins. Indeed, many theories with fundamental Lorentz violation may face fine-tuning problems in order to avoid low-energy Lorentz-violating effects that are several orders of magnitude greater than existing experimental constraints \cite{Collins:2004bp}. However, even a theory which possesses Lorentz invariance at high energies could spontaneously break it at low energies, and with safer experimental consequences.

Spontaneous violation of Lorentz invariance will generally result when a field that transforms non-trivially under the Lorentz group acquires a vacuum expectation value (VEV). A simple example is that of a vector field whose VEV is non-vanishing everywhere. As mentioned above, in order to avoid the experimental constraints such a vector field should not be coupled to the standard model fields, but in order to not be completely innocuous we would like it to couple to gravity. Moreover, to model Lorentz violation in gravity without abandoning the successes of general relativity --- in particular, without giving up general covariance --- the (spontaneously) Lorentz-violating field must be dynamical.

A particularly simple, yet quite general, example of a model with these features is Einstein-aether theory (\ae -theory) \cite{Jacobson:2000xp,Jacobson:2008aj}. It adds to general relativity a dynamical, constant-length timelike vector field, called the aether and denoted by $u^{a}$, which spontaneously breaks Lorentz invariance by picking out a preferred frame at each point in spacetime while maintaining local rotational symmetry (breaking only the boost sector of the Lorentz symmetry) \cite{Eling:2004dk,Jacobson:2008aj}. The constant-length constraint renders non-dynamical a length-stretching mode with a wrong-sign kinetic term \cite{Elliott:2005va}, while also ensuring that the aether picks a globally non-zero VEV and so breaks Lorentz symmetry everywhere. It has been shown that \ae-theory is the most general effective field theory in which the rotation group is unbroken \cite{ArmendarizPicon:2010mz}, so all the results of this work are applicable to any theory which spontaneously violates boost Lorentz invariance while maintaining rotations.

Recently a generalization of \ae -theory has been considered in which a scalar field, $\phi$, can couple to the aether via its divergence, $\theta\equiv\nabla_{a}u^{a}$ \cite{Donnelly:2010cr,Barrow:2012qy,Sandin:2012gq}. This is of particular interest for cosmology because $\theta$ is related to the local Hubble expansion rate. The aether is forced by symmetry to align with the cosmic rest frame in a spatially homogeneous and isotropic background \cite{Carroll:2004ai,Jacobson:2008aj,Lim:2004js,Carruthers:2010ii}, and purely on geometric grounds we find $\theta=3mH$, with $H=\frac{\dot{a}}{a}$ the cosmic time Hubble parameter and $m$ the constant norm of $u^{a}$. The ability to use the expansion rate so freely in the field equations is a departure from general relativity and other purely metric theories, where $H$ is not a covariant scalar as its definitions are all coordinate-dependent. Thus, this extension of pure \ae -theory opens up the interesting possibility of cosmological dynamics depending directly on the expansion rate in a way that is not allowed by general relativity or many modified gravity theories.

This coupling also allows the aether to affect cosmological dynamics directly. This is not possible in ``pure" \ae -theory, as the aether tracks the dominant matter source and hence can only slow down the expansion, via a rescaling of Newton's constant. If the scalar field coupled to $\theta$ is identified with the inflaton, the aether modifies inflationary dynamics. In a simple case, it adds a driving force which can slow down or speed up inflation \cite{Donnelly:2010cr}. This theory with another simple form of the coupling is also closely related (up to the presence of tranverse spin-1 perturbations) to the $\Theta\mathrm{CDM}$ theory, a dark energy theory in which the small cosmological constant is technically natural \cite{Blas:2011en,Audren:2013dwa}.

The coupling between the aether and scalar is contained in the scalar potential, which is allowed to depend on $\theta$. This is a reasonably general approach to coupling the aether to a scalar field: any terms one can write down which do not fit in this framework would have mass dimension 5 or higher and hence not be power-counting renormalizable. We will perform our analysis with the important assumption that $\phi$ drives a period of slow-roll inflation. Hence we consider this theory to be a fairly general model of Lorentz violation in the inflaton sector.

Our aim is to explore the effects of such a coupling at the level of linear perturbations to a cosmological background, and in particular to find theoretical and observational constraints. For reasonable values of the coupling between the aether and the inflaton, these perturbations are unstable and can destroy the inflationary background. This places a constraint on the coupling which is several orders of magnitude stronger than the existing constraints. If the parameters of the theory are chosen to remove the instability, while satisfying existing constraints on the aether VEV, then the effects of the coupling on observables in the cosmic microwave background will be far below the sensitivity of modern experiments.

The remainder of this paper is organized as follows. In \cref{sec:eom} we review Einstein-aether theory and its coupling to a scalar field through $\theta$. In \cref{sec:flatspace} we discuss the behavior of linearized perturbations of the aether and the scalar around a (non-dynamical) flat background, deriving a stability constraint (previously found by another method in \cite{Donnelly:2010cr}) which provides a useful upper bound on the aether-scalar coupling. In \cref{sec:cosmperts}, we discuss the equations of motion for a homogeneous and isotropic cosmology, and set up the cosmological perturbation theory. In \cref{sec:spin1perts} we examine the spin-1 cosmological perturbations of the aether and metric during a phase of quasi-de Sitter inflation. This provides a clear example of the tachyonic instability, which we explore in some depth. In \cref{sec:spin0perts} we look at the spin-0 perturbations, finding the same instability and calculating the scalar power spectrum. Unusually, isocurvature modes do not appear to first order in a perturbative expansion around the aether norm. We give a worked example in \cref{sec:dj} which elucidates the arguments made for a general potential in the preceding sections, and conclude with a discussion of our results in \cref{sec:discussion}.

\section{Einstein-Aether Theory}

\label{sec:eom}

\subsection{Pure Aether Theory}

Einstein-aether theory (which we will often refer to as ``pure" Einstein-aether theory or \ae-theory) is a theory of the spacetime metric $g_{\mu\nu}$ and a vector field (the ``aether") $u^{\mu}$. It is the most general effective theory of Lorentz violation which preserves invariance under rotations \cite{ArmendarizPicon:2010mz}. The action is \cite{Jacobson:2008aj,Carroll:2004ai}
\begin{equation}
S=\int d^{4}x\sqrt{-g}\left[  \frac{1}{16\pi G}R-K^{\mu\nu}{}_{\rho\sigma
}\nabla_{\mu}u^{\rho}\nabla_{\nu}u^{\sigma}+\lambda\left(  u^{\mu}u_{\mu
}+m^{2}\right)  \right]  ,\label{eq:aeaction}
\end{equation}
where
\begin{equation}
K^{\mu\nu}{}_{\rho\sigma}\equiv c_{1}g^{\mu\nu}g_{\rho\sigma}+c_{2}
\delta_{\rho}^{\mu}\delta_{\sigma}^{\nu}+c_{3}\delta_{\sigma}^{\mu}
\delta_{\rho}^{\nu}+c_{4}u^{\mu}u^{\nu}g_{\rho\sigma}.
\end{equation}
The action (\ref{eq:aeaction}) contains an Einstein-Hilbert term for the metric, a kinetic term
for the aether with four dimensionless coefficients $c_{i}$ (coupling the
aether to the metric through the covariant derivatives), and a non-dynamical
Lagrange multiplier $\lambda$. Varying this action with
respect to $\lambda$ constrains the aether to be timelike with a constant norm,
$u^{\mu}u_{\mu}=-m^{2}$.

The action (\ref{eq:aeaction}) is the most general diffeomorphism-invariant
action containing the metric, aether, and up to second derivatives of each.
Most terms one can write down are eliminated by the fixed norm condition,
and other terms such as $R_{\mu\nu}u^{\mu}u^{\nu}$ are equivalent to terms in
\cref{eq:aeaction} under integration by parts. In what follows we will follow
much of the literature on aether cosmology (e.g., \cite{Carroll:2004ai,Lim:2004js}) and
ignore the quartic self-interaction term by setting $c_{4}=0$.

It is generally assumed that (standard-model) matter fields couple to the metric only. Any coupling to the aether
would lead to Lorentz violation in the matter sector by inducing different maximum propagation speeds for different fields, an effect which is strongly constrained by experiment \cite{Mattingly:2005re}. As we are primarily interested in
exploring and constraining Lorentz violation in the gravitational sector and in a single non-standard model scalar, we need not worry about such a coupling. These problematic standard-model couplings may, however, be forbidden by a supersymmetric extension of \ae-theory \cite{Pujolas:2011sk}.

The gravitational constant $G$ that appears in \cref{eq:aeaction} is to be
distinguished from the gravitational constants which appear in the Newtonian
limit and in the Friedmann equations, both of which are modified by the
presence of the aether \cite{Carroll:2004ai}. The Newtonian gravitational
constant, $G_{N}$, and cosmological gravitational constant, $G_{C}$, are
related to the bare constant $G$ by
\begin{align}
G_{N} &  =\frac{G}{1+8\pi G\delta},\\
G_{c} &  =\frac{G}{1+8\pi G\alpha},
\end{align}
where
\begin{align}
\delta &  \equiv-c_{1}m^{2},\\
\alpha &  \equiv(c_{13}+3c_{2})m^{2}. \label{eq:alphadef}
\end{align}
We have introduced the notation $c_{13}\equiv c_{1}+c_{3}$, etc., which we
will use throughout.

\subsection{Coupling to a Scalar Inflaton}

We now introduce to the theory a canonical scalar field $\phi$ which is
allowed to couple kinetically to the aether through its expansion, $\theta\equiv\nabla_{\mu}u^{\mu}$ \cite{Donnelly:2010cr}. The full action reads
\begin{equation}
S=\int d^{4}x\sqrt{-g}\left[  \frac{1}{16\pi G}R - K^{\mu\nu}{}_{\rho\sigma
}\nabla_{\mu}u^{\rho}\nabla_{\nu}u^{\sigma}+\lambda\left(  u^{\mu}u_{\mu
}+m^{2}\right)  -\frac{1}{2}(\partial\phi)^{2}-V(\theta,\phi)\right]
.\label{eq:fullaction}
\end{equation}

Let us pause to motivate the generality of this model. The aim of this paper is to constrain couplings between a Lorentz-violating field and a scalar, particularly a canonical, slowly-rolling scalar inflaton, in as general a way as possible. The model of Lorentz violation is quite general: Einstein-aether theory is the unique Lorentz-violating effective field theory in which rotational invariance is maintained \cite{ArmendarizPicon:2010mz} (although we note that there is an allowed term, the quartic self-interaction parameterized by $c_4$, which we have turned off). Hence, any theory which spontaneously violates Lorentz symmetry without breaking rotational invariance will be described by the vector-tensor sector of our model at low energies.

As for the scalar sector, we have assumed a canonical kinetic term. Moreover, it is clear that there are coupling terms between the aether and the scalar which do not fall under the form $V(\theta,\phi)$. It would be difficult, and is beyond the scope of this paper, to consider such couplings in full generality. However, this form does capture all terms up to mass dimension 4, which one might consider to be dominant, for example, for power-counting renormalizability. This is because the aether, scalar, and derivative operators all have mass dimension 1, the aether norm is constant and cannot be used in the coupling, and because the aether and derivative operators carry spacetime indices which need to be contracted. Any allowed terms involving both $u^\mu$ and $\phi$ up to mass dimension 4 which are not of the form $V(\theta,\phi)$ can be recast into such a form under integration by parts.

Note that in a homogeneous and isotropic background, the aether aligns with
the cosmic rest frame so $\theta$ is essentially just the volume Hubble
parameter, $\theta=3mH$. Hence, the introduction of the aether allows a scalar
inflaton to couple directly to the expansion rate. This is impossible in GR
where $H$ is not proportional to any Lorentz scalar. This was the physical motivation for introducing this type of coupling in \cite{Donnelly:2010cr}.

The aether equation of motion, obtained by varying the action with respect to
$u^{\mu}$, is
\begin{equation}
\lambda u^{\nu}= \nabla_{\mu}J^{\mu\nu}-\frac{1}{2}\nabla^{\nu}V_{\theta} \label{eq:aethereom}
\end{equation}
where the current tensor is defined by
\begin{equation}
J^{\mu}{}_{\sigma}\equiv -K^{\mu\nu}{}_{\sigma\rho}\nabla_{\nu}u^{\rho}.
\end{equation}
Projecting this equation along $u^{\mu}$ allows us to obtain the Lagrange
multiplier $\lambda$,
\begin{equation}
\lambda= -\frac{1}{m^{2}} u_{\nu}\nabla_{\mu}J^{\mu\nu} + \frac{1}{2m^{2}
}u^{\mu}\nabla_{\mu}V_{\theta}.
\end{equation}

The stress-energy tensor for the combined aether-scalar system, taking into
account the contribution from the Lagrange multiplier term, is
\begin{equation}
T_{\mu\nu} = 2\frac{\delta\mathcal{L}}{\delta g^{\mu\nu}} + u^{\rho}
\frac{\delta\mathcal{L}}{\delta u^{\rho}}u_{\mu}u_{\nu}- \mathcal{L}g_{\mu\nu}
\end{equation}
where $\mathcal{L}$ is the Lagrangian for the aether and scalar. Using this
formula we find the stress-energy tensor,
\begin{align}
T_{\mu\nu} ={} &  2c_{1}(\nabla_{\mu}u^{\rho}\nabla_{\nu}u_{\rho}-
\nabla^{\rho}u_{\mu}\nabla_{\rho}u_{\nu})\nonumber\\
&  - 2[\nabla_{\rho}(u_{(\mu} J^{\rho}{}_{\nu)}) + \nabla_{\rho}(u^{\rho
}J_{(\mu\nu)}) - \nabla_{\rho}(u_{(\mu}J_{\nu)}{}^{\rho})]\nonumber\\
&  - 2m^{-2}u_{\sigma}\nabla_{\rho}J^{\sigma\rho} u_{\mu}u_{\nu}+ g_{\mu\nu
}\mathcal{L}_{u}\nonumber\\
&  + \nabla_{\mu}\phi\nabla_{\nu}\phi-\left( \frac{1}{2}\nabla_{\rho}
\phi\nabla^{\rho}\phi+V-\theta V_{\theta}\right) g_{\mu\nu}\nonumber\\
&  +(u^{\rho}\nabla_{\rho}V_{\theta})(g_{\mu\nu}+m^{-2}u_{\mu}u_{\nu
})\label{eq:fullsetensor}
\end{align}	
where $\mathcal{L}_{u} \equiv K^{\mu\nu}{}_{\rho\sigma}\nabla_{\mu}u^{\rho
}\nabla_{\nu}u^{\sigma}$ is the Einstein-aether Lagrangian.

Finally, the inflaton obeys the usual Klein-Gordon equation,
\begin{equation}
\Box\phi=V_{\phi},
\end{equation}
though this is coupled to the aether since generally we will have $V_{\phi}=V_{\phi}(\theta,\phi)$.

\section{Stability Constraint in Flat Spacetime}

\label{sec:flatspace}

Before moving on to the main focus of this paper, perturbations around a
cosmological background, we briefly examine perturbation theory in flat
spacetime. Our goal is to derive a constraint on the coupling $V_{\theta\phi}$
by requiring the aether and scalar perturbations be stable around a Minkowski background.
This will set an upper limit relating the coupling to the effective mass of the scalar,
\begin{equation}
V_{\theta\phi}^{2}(0,0)\leq2c_{123}V_{\phi\phi}(0,0),
\end{equation}
which we will find useful when we examine the cosmological perturbations.

We assume that the potential is analytic around $(\theta,\phi)=(0,0)$, because
if it diverges there the aether-scalar stress-energy tensor
(\ref{eq:fullsetensor}) will be non-zero and we
cannot have a flat spacetime solution. We will also assume that $V(0,0)$ is
either vanishing or negligibly small; if not, then this contributes a
cosmological constant term to the stress-energy tensor, and our background is (anti-)de Sitter rather than
flat. Observations constrain such a term, barring a non-linear screening
mechanism, to be very small.\footnote{The scalar field is canonical, coupled minimally to gravity, and not coupled at all to the matter sector, so we would not expect any screening mechanisms to be present in this theory.}

In flat spacetime the field equations are solved by a constant-field configuration,
\begin{align}
\bar{u}^{\mu} &  =(m,0,0,0),\\
\bar{\lambda} &  =0,\\
\bar{\phi} &  =0.
\end{align}
We introduce small perturbations $\{v^{\mu},\delta\lambda,\delta\phi\}$
defined by
\begin{align}
u^{\mu} &  =\bar{u}^{\mu}+v^{\mu},\\
\lambda &  =\bar{\lambda}+\delta\lambda,\\
\phi &  =\bar{\phi}+\delta\phi.
\end{align}
Writing the action (\ref{eq:fullaction}) as
\begin{equation}
S=\int d^{4}x\mathcal{L,}
\end{equation}
we expand the Lagrangian to quadratic order,
\begin{equation}
\mathcal{L}=\bar{\mathcal{L}}+\delta_{1}\mathcal{L}+\delta_{2}\mathcal{L},
\end{equation}
where $\delta_{1}\mathcal{L}$ and $\delta_{2}\mathcal{L}$ are of linear and
quadratic order, respectively. The background and linear Lagrangians recover
the background equations of motion, leaving us with the quadratic Lagrangian,
\begin{align}
\delta_{2}\mathcal{L}={} &  -c_{1}\partial_{\mu}v^{\nu}\partial^{\mu}v_{\nu
}-c_{2}(\partial_{\mu}v^{\mu})^{2}-c_{3}\partial_{\mu}v^{\nu}\partial_{\nu
}v_{\mu}+2\delta\lambda\left(  \bar{u}^{\mu}v_{\mu}\right) \nonumber\\
&  -\frac{1}{2}\partial_{\mu}\delta\phi\partial^{\mu}\delta\phi-\frac{1}
{2}\left[  V_{\theta\theta}(0,0)(\partial_{\mu}v^{\mu})^{2}+V_{\phi\phi
}(0,0)\delta\phi^{2}+2V_{\theta\phi}(0,0)\delta\phi(\partial_{\mu}v^{\mu
})\right]  ,\label{eq:quadlagfull}
\end{align}
whose variation yields the equations of motion of the perturbed variables.
From here we drop the $(0,0)$ evaluation on the derivatives of the potential
(although they remain implicit). The $\delta\lambda$ equation of motion is
\begin{equation}
\bar{u}^{\mu}v_{\mu}=0.
\end{equation}
It constrains the timelike component of the aether perturbation to vanish:
\begin{equation}
v^{0}=0.
\end{equation}
Inserting this result into \cref{eq:quadlagfull} and splitting $v^{i}$ into
spin-0 and spin-1 fields\footnote{The aether perturbation is in a reducible
subgroup of SO(3), so by decomposing $v^{i}$ like this we single out the real
dynamical degrees of freedom.} as
\begin{equation}
v^{i}=S^{i}+N^{i},
\end{equation}
where the spin-0 piece $S^{i}$ is the divergence of a scalar potential
($S^{i}=\partial^{i}\mathcal{V}$) and the spin-1 piece $N^{i}$ is transverse
to $S^{i}$ ($\partial_{i}N^{i}=0$), we find that the quadratic potential
decouples for these two pieces:
\begin{equation}
\delta_{2}\mathcal{L}=\mathcal{L}^{(0)}+\mathcal{L}^{(1)},
\end{equation}
where the spin-0 Lagrangian is
\begin{align}
\mathcal{L}^{(0)}={} &  c_{1}\dot{S}^{2}-c_{1}\partial_{i}S^{j}\partial
^{i}S_{j}-c_{2}(\partial_{i}S^{i})^{2}-c_{3}\partial_{i}S^{j}\partial_{j}
S^{i}\nonumber\\
&  +\frac{1}{2}\left(  \dot{\delta\phi}^{2}-\delta^{ij}\partial_{i}\delta
\phi\partial_{j}\delta\phi\right)  -\frac{1}{2}\left[  V_{\theta\theta
}(\partial_{i}S^{i})^{2}+V_{\phi\phi}\delta\phi^{2}+2V_{\theta\phi}\delta
\phi(\partial_{i}S^{i})\right] \label{eq:spin0lag}
\end{align}
and the spin-1 Lagrangian is
\begin{equation}
\mathcal{L}^{(1)}=c_{1}\dot{N}^{2}-c_{1}\partial_{i}N^{j}\partial^{i}
N_{j}.\label{eq:spin1lag}
\end{equation}
We have eliminated the cross-terms between the spin-0 and spin-1 pieces, and
the $c_{3}$ term in the spin-1 piece, using integration by parts.

Notice that a consequence of the spin-1 perturbation $N^{i}$ being
divergence-free is that the scalar-field coupling does not affect the spin-1
Lagrangian, because $\phi$ only couples to the aether through $\theta
=\nabla_{\mu}u^{\mu}$. In particular, this allows us to use the constraint
\begin{equation}
c_{1}>0
\end{equation}
from the start. This was derived in pure \ae -theory from requiring positivity
of the quantum Hamiltonian for both the spin-0 and spin-1 fields
\cite{Lim:2004js}, and is suggested by the fact that for $c_{1}\leq0$ the
kinetic terms for $S^{i}$ and $N^{i}$ in \cref{eq:spin0lag,eq:spin1lag} are of
the wrong sign. Since this was proven to be true for the spin-1 perturbations
in \ae -theory and they remain unchanged in this extension of it, this
condition on $c_{1}$ must continue to hold.

Finally, we can vary the action with respect to our three perturbation
variables --- $S^{i}$, $N^{i}$, and $\delta\phi$ --- to obtain the equations
of motion:
\begin{align}
\ddot{S}^{i}-\frac{c_{123}+\frac{1}{2}V_{\theta\theta}}{c_{1}}\partial
^{2}S^{i}-\frac{1}{2c_{1}}V_{\theta\phi}\delta^{ij}\partial_{j}\delta\phi &
=0,\\
\ddot{N}^{i}-\partial^{2}N^{i} &  =0,\\
\Box\delta\phi-V_{\phi\phi}\delta\phi-V_{\theta\phi}\partial_{i}S^{i} &  =0.
\end{align}
In \ae -theory, $\phi=0=V(\theta,\phi)$ and both aether equations are simply
wave equations with plane wave solutions \cite{Lim:2004js},
\begin{align}
S^{i}(\vec{k}) &  \propto e^{-ic_{s}^{(0)}kt+i\vec{k}\cdot\vec{x}
},\label{eq:spin0flat}\\
N^{i}(\vec{k}) &  \propto e^{-ic_{s}^{(1)}kt+i\vec{k}\cdot\vec{x}
},\label{eq:spin1flat}
\end{align}
with the propagation speeds for the spin-0 and spin-1 perturbations given by
\begin{align}
c_{s}^{(0)2} &  =\frac{c_{123}}{c_{1}},\\
c_{s}^{(1)2} &  =1.
\end{align}

The scalar coupling modifies the \ae-theory situation in two ways. First, $c_{123}$ is
shifted by $\frac{1}{2}V_{\theta\theta}$ evaluated at $(\theta=0,\phi=0)$
(remember that implicitly we are evaluating all the derivatives of $V$ there,
so they are just constants). This is to be expected: the expansion of the
potential around $(0,0)$ includes, at second order, the term $\frac{1}
{2}V_{\theta\theta}\delta\theta^{2}=\frac{1}{2}V_{\theta\theta}(\partial
_{i}S^{i})^{2}$, which can be absorbed into the $c_{2}$ term in the
(quadratic) Lagrangian by redefining $c_{2}\rightarrow c_{2}+\frac{1}
{2}V_{\theta\theta}$. We will find this same redefinition of $c_{2}$ appears
in the cosmological perturbation theory.

The second change from \ae -theory is more significant for the dynamics. When
$V_{\theta\phi}$ is non-zero --- i.e., when the coupling between $u^{\mu}$ and
$\phi$ is turned on --- it adds a source term to the wave equation for $S^{i}$
(the $N^{i}$ equation is unmodified because neither $\theta$ nor $\phi$
contain spin-1 pieces, as discussed above). Similarly, a $u^{\mu}$-dependent
source is added to the quadratic order Klein-Gordon equation for $\delta\phi$.

The equations of motion for $S^i$ and $\delta\phi$ are those of two coupled harmonic oscillators. To simplify these, we move to Fourier space, where the spin-0
degrees of freedom $S_{k}^{i}(t)=\partial^{i}\mathcal{V}_{k}(t)$ and
$\delta\phi_{k}(t)$ obey the coupled wave equations (dropping the $k$
subscripts and absorbing $\frac{1}{2}V_{\theta\theta}$ into $c_{2}$):
\begin{align}
\ddot{\mathcal{V}}+c_{s}^{(0)2}k^{2}\mathcal{V}-\frac{1}{2c_{1}}V_{\theta\phi
}\delta\phi &  =0,\\
\ddot{\delta\phi}+(k^{2}+V_{\phi\phi})\delta\phi-V_{\theta\phi}k^{2}
\mathcal{V} &  =0.
\end{align}
This system can be diagonalized\footnote{We thank the referee for this suggestion, which simplifies a calculation done in an earlier draft while obtaining the same result.} by defining
\begin{align}
\tilde{\mathcal{V}} &\equiv \mathcal{V} + \frac{V_{\theta\phi}}{2c_1\left(k^2+V_{\phi\phi}^2-\omega^2_-\right)}\delta\phi, \\
\tilde{\delta\phi} &\equiv \delta\phi + \frac{V_{\theta\phi} k^2}{c_s^{(0)2}k^2 - \omega^2_+}\mathcal{V},
\end{align}
where the $\omega_\pm$ are defined by
\begin{equation}
2\omega^2_\pm \equiv k^2\left(1+c_s^{(0)2}\right) + V_{\phi\phi}^2 \pm \sqrt{\left[k^2 \left(1-c_s^{(0)2}\right) + V_{\phi\phi}^2\right]^2 + \frac{2V_{\theta\phi}^2k^2}{c_1}}.
\end{equation}
Under this transformation, the equations of motion are simply
\begin{align}
\ddot{\tilde{\mathcal{V}}} + \omega^2_-\tilde{\mathcal{V}} &=0, \label{eq:tildev} \\
\ddot{\tilde{\delta\phi}} + \omega^2_+\tilde{\delta\phi} &=0. \label{eq:tildedphi}
\end{align}
Note that in the limit $V_{\theta\phi} \to 0$ where the two fields decouple, $\omega^2_+$ goes to $k^2+V_{\phi\phi}^2$, the squared frequency of a $\delta\phi$ mode, and $\omega^2_-$ goes to $c_s^2k^2$, the equivalent for $\mathcal{V}$ modes. We see that $\tilde{\mathcal{V}}$ and $\tilde{\delta\phi}$ are non-interacting, mixed modes which reduce to $\mathcal{V}$ and $\delta\phi$, respectively, in the absence of a scalar-aether coupling.

For stability, we require the $\omega_\pm$ to be real, so that the solutions to \cref{eq:tildev,eq:tildedphi} are plane waves
rather than growing and decaying exponentials. Note that $\omega_+$ is manifestly real, so the $\tilde{\delta\phi}$ modes are always stable. It is the aether modes, $\tilde{\mathcal{V}}$, which can be destabilized by the coupling to the scalar, while the reverse is not true. Stability imposes a constraint on $V_{\theta\phi}$,
\begin{equation}
V_{\theta\phi}^{2}\leq2c_{1}c_{s}^{(0)2}(k^{2}+V_{\phi\phi}),
\end{equation}
which, since we would like it to hold for arbitrarily large wavelength modes
(small $k$), can be written (substituting back in the definition $c_{s}
^{(0)2}=c_{123}/c_{1}$)
\begin{equation}
V_{\theta\phi}^{2}(0,0)\leq2c_{123}V_{\phi\phi}
(0,0),\label{eq:spin0constraint}
\end{equation}
where for clarity we have put back in the $(\theta,\phi)=(0,0)$ evaluation
which has been implicit. \Cref{eq:spin0constraint} constrains the coupling
between the aether and the scalar field in terms of the aether kinetic free parameters (or, equivalently, its no-coupling
propagation speed) and the effective mass of the scalar in flat spacetime. It agrees with the
spin-0 stability constraint in \cite{Donnelly:2010cr} which was derived in a
slightly different fashion for a specific form of $V(\theta,\phi
).$\footnote{Our notation is different than that used in \cite{Donnelly:2010cr}
and as a result their constraint looks slightly different. They define the aether to be dimensionless (and unit norm) while we give it a norm $m$ with mass dimensions. To compensate for this, their $c_i$ are $16\pi Gm^2c_i$ in our notation. We have checked, translating between the two notations, that our constraint matches theirs.} The $c_i$ are dimensionless, so we might expect them to naturally be $\mathcal{O}(1).$ Assuming this, \cref{eq:spin0constraint} roughly constrains the coupling $V_{\theta\phi}(0,0)$ to be less than the scalar field mass around flat spacetime. Note that this constraint also implies
$c_{123}\geq0$,\footnote{Assuming that the scalar field is non-tachyonic.}
which is the combined constraint from subluminal propagation and positivity of
the Hamiltonian of the spin-0 field in pure \ae -theory \cite{Lim:2004js}.

\section{Cosmological Perturbation Theory}
\label{sec:cosmperts}

The goal of this paper is to explore the impact of the coupling between $\phi$ and
$u^{\mu}$ on small perturbations to a homogeneous and isotropic
cosmology. We will be particularly interested in a period of slow-roll
inflation driven by $\phi$. As has been explored in great depth over the past
three decades, a scalar field rolling slowly down its potential can lead to
cosmic inflation and all of the interesting cosmological consequences for
explaining the structure of the observed universe that follow from it
\cite{PeterUzan}. In this section we present the metric
and scalar field equations in a homogeneous and isotropic universe and set up the cosmological perturbations.

\subsection{Background Cosmology}

We restrict to a flat Friedmann-Robertson-Walker background geometry evolving
in conformal time, $\tau$,
\begin{equation}
ds^{2}=a^{2}(\tau)(-d\tau^{2}+d\vec{x}^{2}).\label{eq:FRW}
\end{equation}
The $0-0$ and trace Einstein equations give us the Friedmann equations:
\begin{align}
\mathcal{H}^{2} &  =\frac{8\pi G_{c}}{3}a^{2}\left(  V-\theta V_{\theta}
+\rho_{m}+\frac{1}{2}\phi^{\prime2}a^{-2}\right) \label{eq:friedmann} ,\\
\mathcal{H}^{\prime} &  =\frac{4\pi G_{c}}{3}a^{2}\left[  -3\frac{m}{a}\left(
3\frac{m}{a}V_{\theta\theta}(\mathcal{H}^{\prime}-\mathcal{H}^{2}
)+V_{\theta\phi}\phi^{\prime}\right)  -\rho_{m}(1+3w)+2(V-\theta V_{\theta
})-2\phi^{\prime2}a^{-2}\right]  , \label{eq:friedmann2}
\end{align}
where $\mathcal{H}\equiv a^{\prime}/a= d\ln a/d\tau$ is the
conformal time Hubble parameter, and, as discussed in \Cref{sec:eom}, the
effective cosmological gravitational constant, $G_{c}$, is related to the bare
constant, $G$, by
\begin{equation}
G_{c}=\frac{G}{1+8\pi G\alpha},
\end{equation}
with $\alpha=(c_{1}+3c_{2}+c_{3})m^{2}$. This modification of Newton's
constant arises because in a homogeneous and isotropic background the
Einstein-aether terms for the vector field only contribute stress-energy that
tracks the dominant matter fluid, so the associated energy density is
proportional to $H^{2}$ \cite{Carroll:2004ai}. The only dynamical stress-energy from the aether, in
the background, is that due to the scalar coupling. However, the aether
perturbations do carry some dynamics even in the absence of the coupling to $\phi$ \cite{Lim:2004js}. For completeness we have included a matter component with equation of state $w$, although from now on we will assume that $\phi$ is gravitationally dominant and ignore any $\rho_m$.

The scalar field obeys the usual cosmological Klein-Gordon equation:
\begin{equation}
\phi^{\prime\prime}+2{\mathcal{H}}\phi^{\prime} + a^2V_{\phi}=0.
\end{equation}
The coupling to $\theta$ is contained in the function $V_{\phi}$. In
the background, $\theta=3mH$, with $H=\mathcal{H}/a$ the cosmic time Hubble
parameter, so this contributes extra Hubble friction or driving
\cite{Donnelly:2010cr}.

We need not write down the aether field equations, at least in the background.
The vector field must be aligned with the cosmic rest frame due to homogeneity
and isotropy, and its value --- $u^{\mu}=ma^{-1}\delta^{\mu}{}_{0}$ --- is
determined completely by the normalization condition $u_{\mu}u^{\mu}=-m^{2}$.
One can check that this solution satisfies the spatial component of the aether
equation, while the temporal component only determines the Lagrange
multiplier. In pure \ae-theory this solution is stable perturbatively
\cite{Lim:2004js,Kanno:2006ty,Li:2007vz,ArmendarizPicon:2010rs} and that
stability holds nonlinearly for most large perturbations \cite{Carruthers:2010ii}. This statement is subject to several constraints on the $c_i$ parameters which can be found in, e.g., \cite{Lim:2004js,Jacobson:2008aj,ArmendarizPicon:2010rs}, and we will assume throughout our analysis that these constraints hold. One of the important results of this paper is that the coupling between $u^\mu$ and $\phi$ can render this solution \textit{unstable} for large regions of parameter space that are allowed by other experimental, observational, and theoretical constraints.

When the scalar potential is $V(\theta,\phi)=V(\phi)$, the background aether is irrelevant apart from rescaling Newton's constant, and many choices for the potential can lead to periods
of slow-roll inflation \cite{LythLiddle}. Adding a coupling to the aether
will change the dynamics but may still allow for inflation \cite{Donnelly:2010cr}. We will therefore aim to be as general about $V(\theta,\phi)$ as possible when discussing perturbation theory.

\subsection{Perturbation Variables}

Let us consider linear perturbations about the FRW background (\ref{eq:FRW}).
We will work with the perturbed metric
\begin{equation}
ds^{2}=a^{2}(\eta)\left\{  -(1+2\Phi)d\eta^{2}-2B_{i}d\eta dx^{i}
+[(1+2\Psi)\delta_{ij}+2H_{Tij}]dx^{i}dx^{j}\right\}  ,
\end{equation}
so the perturbed metric components are
\begin{align}
g_{00} &  =-a^{2}(1+2\Phi),\nonumber\\
g_{0i} &  =-a^{2}B_{i},\nonumber\\
g_{ij} &  =a^{2}[(1+2\Psi)\delta_{ij}+2H_{Tij}].\nonumber
\end{align}
Inverting, and keeping terms to first-order we have
\begin{align}
g^{00} &  =-a^{-2}(1-2\Phi),\nonumber\\
g^{0i} &  =-a^{-2}B^{i},\nonumber\\
g^{ij} &  =a^{-2}[(1-2\Psi)\delta^{ij}-2H_{T}^{ij}].\nonumber
\end{align}
Indices on spatial quantities like $B_{i}$ are raised and lowered with
$\delta_{ij}$. The Christoffel symbols are (with background parts in bold)
\begin{align}
\Gamma_{00}^{0} &  =\boldsymbol{\mathcal{H}}+\Phi^{\prime},\nonumber\\
\Gamma_{0i}^{0} &  =\Phi_{,i}-\mathcal{H}B_{i},\nonumber\\
\Gamma_{ij}^{0} &  =\left(  \boldsymbol{\mathcal{H}}(1+2\Psi)+\Psi^{\prime
}-2\mathcal{H}\Phi\right)  \delta_{ij}+B_{(i,j)}+(2\mathcal{H}H_{Tij}
+H_{Tij}^{\prime}),\nonumber\\
\Gamma_{00}^{i} &  =\Phi^{,i}-\mathcal{H}B^{i}-B^{\prime i},\nonumber\\
\Gamma_{0j}^{i} &  =\boldsymbol{\mathcal{H}\delta_{j}^{i}}+\delta
^{ik}B_{[j,k]}+\Psi^{\prime}\delta_{j}^{i}+H_{T}^{\prime}{}_{j}^{i}
,\nonumber\\
\Gamma_{jk}^{i} &  =\mathcal{H}B^{i}\delta_{jk}+\Psi_{,k}\delta_{j}^{i}
+\Psi_{,j}\delta_{k}^{i}-\Psi^{,i}\delta_{jk}+H_{T}{}_{j,k}^{i}+H_{T}{}
_{k,j}^{i}-H_{Tjk}{}^{,i}.\nonumber
\end{align}
We do not reproduce the Einstein tensor components here; they can be found in the
literature \cite{Kodama:1985bj}.

The aether in the background has only $u^{0}=\frac{m}{a}$. Imposing the
constant norm constraint, $u_{\mu}u^{\mu}=-m^{2}$, to first order the aether is
\begin{equation}
u^{\mu}=\frac{m}{a}\left(  (1-\Phi),V^{i}\right)  ,\label{eq:perturbedaether}
\end{equation}
and with lowered indices we have
\begin{equation}
u_{\mu}=ma\left(  -(1+\Phi),V_{i}-B_{i}\right),
\end{equation}
where spatial indices on $V^{i}$ and $B^{i}$ are raised and lowered using the
spatial metric $\delta_{ij}$.
Taking the divergence of \cref{eq:perturbedaether} we find the perturbed
expansion to be
\begin{equation}
\theta=\frac{m}{a}\left[  3\mathcal{H}(1-\Phi)+3\Psi^{\prime}+V^{i}{}
_{,i}\right]  .
\end{equation}

The scalar field $\phi$ is split into a background piece and a small
perturbation,
\begin{equation}
\phi= \bar\phi+ \delta\phi,
\end{equation}
where $\bar\phi$ satisfies the Klein-Gordon equation in the background metric.

\subsection{Perturbed Equations of Motion}

In deriving the perturbation equations we will need to expand $V(\theta
.\phi)$ around its background value:
\begin{equation}
V(\theta,\phi)=\bar{V}+\bar{V}_{\theta}\delta\theta+\bar{V}_{\phi}\delta
\phi+\frac{1}{2}\left[  \bar{V}_{\theta\theta}\delta\theta^{2}+\bar{V}
_{\phi\phi}\delta\phi^{2}+2\bar{V}_{\theta\phi}\delta\theta\delta\phi\right]
+O(\delta\theta^{3}),
\end{equation}
where
\begin{equation}\delta\theta  =\frac{m}{a}\left(  3\Psi^{\prime}-3\mathcal{H}\Phi+V^{i}
{}_{,i}\right),
\end{equation}
so the relevant expansions are
\begin{align}
V(\theta,\phi) &  =\bar{V}+\bar{V}_{\theta}\delta\theta+\bar{V}_{\phi}
\delta\phi+O(\delta\theta^{2}),\nonumber\\
V_{\theta}(\theta,\phi) &  =\bar{V}_{\theta}+\bar{V}_{\theta\theta}
\delta\theta+\bar{V}_{\theta\phi}\delta\phi+O(\delta\theta^{2}),\nonumber\\
V_{\phi}(\theta,\phi) &  =\bar{V}_{\phi}+\bar{V}_{\phi\phi}\delta\phi+\bar
{V}_{\theta\phi}\delta\theta+O(\delta\theta^{2}),\nonumber\\
V_{\theta\theta}(\theta,\phi) &  =\bar{V}_{\theta\theta}+\bar{V}_{\theta
\theta\theta}\delta\theta+\bar{V}_{\theta\theta\phi}\delta\phi+O(\delta
\theta^{2}),\nonumber\\
V_{\theta\phi}(\theta,\phi) &  =\bar{V}_{\theta\phi}+\bar{V}_{\theta\theta
\phi}\delta\theta+\bar{V}_{\theta\phi\phi}\delta\phi+O(\delta\theta)^{2}.
\end{align}
We use overbars throughout this paper to denote background values.

The perturbed equations of motion in real space are given in
\cref{app:realspace}. However, the symmetries of the FRW background allow us
to decompose the perturbations into spin-0, spin-1, and spin-2 components
\cite{Bardeen:1980kt}. In particular, because the background variables (including
the aether, which points only in the time direction) do not break the SO(3)
symmetry on spatial slices, these components conveniently decouple from each other. Hence we perform this
decomposition both to isolate the fundamental degrees of freedom from each
other and to make close contact with the rest of the literature on
cosmological perturbation theory.

We decompose the variables as
\begin{align}
\delta\phi &  =\sum_{k}\delta\phi_{k}Y^{(0)},\\
\Phi &  =\sum_{k}\Phi_{k}Y^{(0)},\\
\Psi &  =\sum_{k}\Psi_{k}Y^{(0)},\\
V^{i} &  =\sum_{k}\sum_{m=0,1}V_{k}^{(\pm m)}Y^{i}{}^{(\pm m)},\\
B^{i} &  =\sum_{k}\sum_{m=0,1}B_{k}^{(\pm m)}Y^{i}{}^{(\pm m)},\\
H_{T}^{ij} &  =\sum_{k}\sum_{m=0,1,2}H_{Tk}^{(\pm m)}{}Y^{ij}{}^{(\pm m)},
\end{align}
where the $Y^{(0)}$, etc., are eigenmodes of the Laplace-Beltrami operator
$\partial^{2}+k^{2}$ (see \cite{Kodama:1985bj,Lim:2004js} for the forms of
these mode functions and some of their useful properties). From here on, we will drop
the $k$ subscripts. The spin-0, spin-1, and spin-2 perturbation equations can
then be found by plugging these expansions into the real space equations listed in \cref{app:realspace}.

\section{Spin-1 Cosmological Perturbations}

\label{sec:spin1perts}

We begin our analysis by focusing on the spin-1 perturbations. The spin-2 perturbations
are unmodified by the aether-scalar coupling because $V(\theta,\phi)$ only
contains spin-0 and spin-1 terms. The only physical spin-2 perturbations are the transverse and traceless parts of the metric
perturbation $H_{Tij}$, or gravitational waves, and they behave as they do in pure
\ae -theory \cite{Lim:2004js}. The spin-0 perturbations, discussed in \cref{sec:spin0perts}, are more complicated than the spin-1 perturbations due to the presence of $\delta\phi$ modes. The important physical result --- the existence of unstable perturbations for large, otherwise experimentally-allowed regions of parameter space --- will therefore be easier to see and understand in the context of the simpler spin-1 modes.

The only non-trivial spin-1 component of the aether field equation is $\nu=i
$,
\begin{align}
&  \left\{  \left[  -2\frac{\alpha}{m^{2}}{\mathcal{H}}^{2}+\frac{\alpha
}{m^{2}}\frac{a^{\prime\prime}}{a}-c_{1}\frac{a^{\prime\prime}}{a}\right]
(B^{(\pm1)}-V^{(\pm1)})\right. \nonumber\\
&  +2c_{1}{\mathcal{H}}(V^{\prime(\pm1)}-B^{\prime(\pm1)})+\frac{1}{2}
(c_{3}-c_{1})k^{2}B^{(\pm1)}+c_{1}k^{2}V^{(\pm1)}\nonumber\\
&  \left.  -c_{13}\frac{k}{2}H_{T}^{\prime}{}^{(\pm1)}-c_{1}(B^{\prime
\prime(\pm1)}-V^{\prime\prime(\pm1)})\right. \nonumber\\
&  +\left.  \frac{1}{2}\left[  3\bar{V}_{\theta\theta}\left(  \frac
{a^{\prime\prime}}{a}-2\mathcal{H}^{2}\right)  +\frac{a}{m}\bar{V}_{\theta
\phi}\bar{\phi}^{\prime}\right]  \left(  B^{(\pm1)}-V^{(\pm1)}\right)
\right\}  Y_{i}^{(\pm1)}=0,\label{eq:spin1vi}
\end{align}
while the spin-1 perturbations of the stress-energy tensor are
\begin{align}
\delta T^{0}{}_{0}^{(\pm1)}={} &  0,\\
\delta T^{0}{}_{i}^{(\pm1)}={} &  \left\{  2\left(  \frac{m}{a}\right)
^{2}\left[  \left(  -2\frac{\alpha}{m^{2}}{\mathcal{H}}^{2}+\frac{\alpha
}{m^{2}}\frac{a^{\prime\prime}}{a}-c_{1}\frac{a^{\prime\prime}}{a}\right)
(V^{(\pm1)}-B^{(\pm1)})\right.  \right. \nonumber\\
&  \left.  -\left.  c_{1}a^{-2}\left[  a^{2}(V^{\prime(\pm1)}-B^{\prime(\pm
1)})\right]  ^{\prime}+\frac{1}{2}(c_{1}-c_{3})k^{2}(B^{(\pm1)}-V^{(\pm
1)})\right]  \right. \nonumber\\
&  +\left.  \frac{m}{a}\left[  \frac{3m}{a}\bar{V}_{\theta\theta}\left(
\frac{a^{\prime\prime}}{a}-2\mathcal{H}^{2}\right)  +\bar{V}_{\theta\phi}
\bar{\phi}^{\prime}\right]  \left(  V^{(\pm1)}-B^{(\pm1)}\right)  \right\}
Y_{i}^{(\pm1)},\label{eq:spin10i}\\
\delta T^{i}{}_{j}^{(\pm1)}={} &  2\left(  \frac{m}{a}\right)  ^{2}
c_{13}\left\{  a^{-2}\left[  a^{2}(-kV^{(\pm1)}+H^{\prime}{}_{T}^{(\pm
1)})\right]  ^{\prime}\right\}  Y^{i}{}_{j}{}^{(\pm1)},
\end{align}
where $\alpha=(c_{13}+3c_2)m^2$ as defined in \cref{eq:alphadef}. As a consistency check, these expressions reduce to those found in the
literature for a scalar field uncoupled to the aether \cite{Mukhanov:1990me}
(setting $V(\theta,\phi)=V(\phi)$) and for \ae -theory \cite{Lim:2004js} (setting $V(\theta
,\phi)=\alpha\theta^{2}$, with $c_{2}\rightarrow c_{2}+\alpha$). For
convenience, from here on we will absorb $\frac{1}{2}\bar{V}_{\theta\theta}$ into $c_{2}$ and indicate
the change with a tilde, e.g.,
\begin{equation}
\tilde{\alpha}\equiv\left(  c_{1}+3c_{2}+c_{3}+\frac{3}{2}\bar{V}
_{\theta\theta}\right)  m^{2}\text{,}
\end{equation}
and similarly for quantities like $\tilde G_{c}$. While this is convenient
notation we should remember that $\bar{V}_{\theta\theta}$ and hence all tilded
quantities are not necessarily constant, although they are nearly so during a
slow-roll phase.\footnote{A non-constant $\bar{V}_{\theta\theta}$ requires
cubic or higher order terms in the potential. For the quadratic Donnelly-Jacobson potential discussed in \cref{sec:dj}, $\bar{V}
_{\theta\theta}$ is constant, and can be freely set to zero by absorbing it
into $c_{2}$.}

We should first note that due to its direct coupling to the aether, the scalar
field \textit{does} source spin-1 perturbations, which is impossible in the
uncoupled case as the scalar field itself contains no spin-1 piece. In pure
\ae -theory the spin-1 perturbations decay away as $a^{-1}$ \cite{Lim:2004js}.
We may wonder if the scalar-vector coupling can counteract this and generate a
non-decaying spin-1 spectrum.

Using the gauge freedom in the spin-1 Einstein equations, we choose to work in
a gauge where $H_{T}^{{(\pm1)}}=0$; that is, we foliate spacetime with
shear-free hypersurfaces. The $i-j$ Einstein equation in the spin-1 case is
unmodified from the \ae -theory case \cite{Lim:2004js} and gives a constraint
relating the shift $B^{{(\pm1)}}$ and the spin-1 aether perturbation
$V^{{(\pm1)}}$,
\begin{equation}
B^{{(\pm1)}}=\gamma V^{{(\pm1)}},
\end{equation}
where
\begin{equation}\gamma\equiv16\pi Gm^{2}c_{13}.
\end{equation}
It is tempting to notice the similarities between the $v=i$ aether field equation (\ref{eq:spin1vi}) and
the $0-i$ Einstein equation (\ref{eq:spin10i}), but this is just hinting at
the underlying redundancy between the two equations. Indeed, using
\cref{eq:spin1vi} to eliminate the scalar field term in \cref{eq:spin10i} just
leaves us with $0=0$. This is because, due to the constraint
equation, the two perturbations $B$ and $V$ are related, and hence (by the
Bianchi identities) these two equations have to contain the same content.
We choose to use the $0-i$ Einstein equation to derive our equation of
motion for the spin-1 perturbations. In this equation, the scalar field
couples to the vector perturbations of the aether and the metric via $\frac
{m}{a}\bar{V}_{\theta\phi}\bar{\phi}^{\prime}$. In the quadratic coupled
potential of Donnelly and Jacobson, discussed in detail in \cref{sec:dj}, the coupling $\bar V_{\theta\phi}$ is
exactly constant. In general, we will take $\bar{V}_{\theta\phi}$ to be
constant to first order in slow roll.

Inserting the constraint into the $0-i$ Einstein equation we find
\begin{align}
&  \left(  2\frac{\alpha}{m^{2}}{\mathcal{H}}^{2}-\frac{\alpha}{m^{2}}
\frac{a^{\prime\prime}}{a}+c_{1}\frac{a^{\prime\prime}}{a}\right)  V^{(\pm
1)}\nonumber\\
&  +\frac{1}{2}\left[  (c_{1}-c_{3})+\frac{c_{13}}{1-\gamma}\right]
k^{2}V^{(\pm1)}+c_{1}(2{\mathcal{H}}V^{(\pm1)\prime}+V^{(\pm1)\prime\prime})\nonumber\\
&  -\frac{1}{2}\left[  3\bar{V}_{\theta\theta}\left(  \frac{a^{\prime\prime}
}{a}-2\mathcal{H}^{2}\right)  +\frac{a}{m}\bar{V}_{\theta\phi}\bar{\phi
}^{\prime}\right]  V=0.\label{eq:spin10iv}
\end{align}
Following \cite{Lim:2004js}, we define $\xi=aV^{{(\pm1)}}$ to eliminate the
first-derivative terms, so \cref{eq:spin10iv} becomes
\begin{equation}
\xi^{\prime\prime}+c_{s}^{{(\pm1)}2}k^{2}\xi+\left(  A\frac{\tilde{\alpha}
}{m^{2}c_{1}}-\frac{1}{2}\frac{a}{mc_{1}}\bar{V}_{\theta\phi}\bar{\phi
}^{\prime}\right)  \xi=0,\label{eq:xieom}
\end{equation}
where the no-coupling sound speed $c_{s}^{{(\pm1)}}$ is the de Sitter propagation speed
of the spin-1 aether and metric perturbations when the coupling to the scalar
field is absent \cite{Lim:2004js},
\begin{equation}
c_{s}^{{(\pm1)}2}=\frac{1}{2}\left[  (1-c_{3}/c_{1})+\frac{1+c_{3}/c_{1}
}{1-\gamma}\right]  ,
\end{equation}
and as before we have absorbed the background $\bar{V}_{\theta\theta}$ into
$c_{2}$ and noted the redefined constants (here, $\alpha$) with a tilde. The
background quantity $A$ is defined by
\begin{equation}
A\equiv2\mathcal{H}^{2}-\frac{a^{\prime\prime}}{a}={\mathcal{H}}
^{2}-{\mathcal{H}}^{\prime}=-a\left(  \frac{{\mathcal{H}}}{a}\right)
^{\prime}
\end{equation}
and vanishes in the de Sitter limit.

\subsection{Slow-Roll Solution}

\label{sec:vecslowroll}

The equation of motion (\ref{eq:xieom}) for the spin-1 aether and metric perturbations, is difficult to solve in full generality. It was solved in
pure de Sitter spacetime ($A=0$) in \ae-theory (i.e., in the absence of the scalar field) in \cite{Lim:2004js}. In that limit,
\cref{eq:xieom} is a wave equation with real frequency, so $\xi$ was found to be
oscillatory. Therefore, in \ae-theory the spin-1 shift perturbation $B^{{(\pm1)}}=\gamma\xi/a$
decays exponentially,\footnote{Here and in the rest of the paper, ``exponential" growth or decay should be taken to mean exponential in cosmic time, or as a power law in conformal time.} leaving the post-inflationary universe devoid of spin-1 perturbations. To investigate whether the inflaton coupling term will change
this conclusion, we solve \cref{eq:xieom} in the slow-roll limit.

We define the slow-roll parameters
\begin{alignat}{2}
\varepsilon &=-\frac{\dot{H}}{H^{2}} &&=1-\frac{{\mathcal{H}}^{\prime}}{{\mathcal{H}}^{2}},\\
\eta &=\phantom{-}\frac{\dot{\varepsilon}}{H\varepsilon} &&=\frac{\varepsilon
^{\prime}}{{\mathcal{H}}\varepsilon}.
\end{alignat}
The slow-roll limit is $\varepsilon,\eta\ll1$. In this limit, both parameters are constant at first order and we can find
\begin{align}
a &  \approx-\frac{1}{H\tau}(1+\varepsilon),\\
{\mathcal{H}} &  \approx-\frac{1}{\tau}(1+\varepsilon
).\label{eq:generalshslowroll}
\end{align}
Taking conformal time derivatives we find
\begin{equation}
A={\mathcal{H}}^{2}-{\mathcal{H}}^{\prime} \approx \frac{\varepsilon}{\tau^{2}}.
\end{equation}

Using these relations, as well as the Klein-Gordon equation in the slow-roll
limit and the fact that ${\mathcal{H}}=aH$, we can write the $\xi$ equation of
motion to first order in slow roll as
\begin{equation}
\xi^{\prime\prime}+c_{s}^{{(\pm1)}2}k^{2}\xi+\frac{1}{\tau^{2}}\left(
\frac{\tilde{\alpha}}{m^{2}c_{1}}\varepsilon+\frac{1}{6mc_{1}}\frac{1+2\varepsilon
}{H^{3}}\bar{V}_{\theta\phi}\bar{V}_{\phi}\right)  \xi
=\mathcal{O}(\varepsilon^2).\label{eq:spin1xislowrollo1}
\end{equation}
$\bar{V}(\theta,\phi)$ and its derivatives will be constant to first order in
the slow-roll parameters, so if we ignore $\mathcal{O}(\varepsilon)$ terms then $\bar{V}_{\phi}$ and $\bar{V}_{\theta\phi}$ in \cref{eq:spin1xislowrollo1} are constants. Our equation of motion for the spin-1 perturbations can then be written simply as
\begin{equation}
\boxed{\xi'' + c_s^{\pmvec2}k^2\xi - \frac{\Lambda}{\tau^2}\xi = 0}\label{eq:spin1xislowroll}
\end{equation}
where we have defined the constant
\begin{equation}
\Lambda \equiv -\frac{\bar{V}_{\phi}\bar{V}_{\theta\phi}}{6mc_{1}H^{3}}+\mathcal{O}(\varepsilon).\label{eq:lambdadef}
\end{equation}
We have also assumed that $\Lambda$ dominates $\tilde{\alpha}\varepsilon
/(m^{2}c_{1})\sim\varepsilon$ (which is $\sim\mathcal{O}(10^{-2})$ \cite{Ade:2013zuv,Ade:2013uln}), the term from pure \ae -theory. In principle this need not be
true, if the coupling term $\bar{V}_{\theta\phi}$ were extraordinarily small. If the aether-scalar
coupling is to do anything interesting, then $\bar{V}_{\theta\phi}$ must be
larger than that, so we will continue to assume that it is.

Notice that the aether perturbation, $V=\xi/a$, has an effective mass in the slow-roll regime,
\begin{equation}
M_\mathrm{eff}^2 = (2-\Lambda) H^2 = 2H^2-\frac{\bar{V}_{\phi}\bar{V}_{\theta\phi}}{6mc_{1}H}\left(1+\mathcal{O}(\varepsilon)\right). \label{eq:meffsq}
\end{equation}
We expect a tachyonic instability for negative $M_\mathrm{eff}^2$, i.e., for $\Lambda>2$. We proceed to demonstrate that just an instability arises.

\subsection{Full Solution}

Noticing the similarity between \cref{eq:spin1xislowroll} and the usual
Mukhanov-Sasaki equation \cite{Mukhanov:1990me}, which has solutions in terms of Bessel functions, we
change variables to $g=x^{-1/2}\xi$ with $x=-c_{s}^{{(\pm1)}}k\tau$ to recast
\cref{eq:spin1xislowroll} as Bessel's equation for $g(x)$,
\begin{equation}
x^{2}\frac{d^{2}g}{dx^{2}}+x\frac{dg}{dx}+\left(  x^{2}-\nu^{2}\right)  g=0,
\end{equation}
with the order $\nu$ given by
\begin{equation}
\nu^{2}\equiv\frac{1}{4}+\Lambda.
\end{equation}
Depending on the sign and magnitude of $\Lambda$, the order $\nu$ can be real
or imaginary. We will find it convenient to write the general solution in
terms of the Hankel functions as
\begin{equation}
\xi=\frac{\sqrt{\pi}}{2}\sqrt{-\tau}\left[  \alpha_{k}H_{\nu}^{(1)}
(-c_{s}^{{(\pm1)}}k\tau)+\beta_{k}H_{\nu}^{(2)}(-c_{s}^{{(\pm1)}}
k\tau)\right]  .
\end{equation}
To determine the values of the Bogoliubov coefficients $\alpha_{k}$ and
$\beta_{k}$ we need to match this solution in the sub-horizon limit
$-c_{s}^{{(\pm1)}}k\tau\rightarrow\infty$ to the quantum vacuum state of the
aether perturbations in flat spacetime. This is desirable because we can
assume that at such short wavelengths, these modes do not \textquotedblleft
see" the cosmic expansion. In section~IV.B of \cite{Lim:2004js} the quantum
mode functions $N_{k}$ for the aether perturbation $v^{i}$ were demonstrated
to satisfy
\begin{equation}
N_{k}=\frac{1}{4\sqrt{|c_{1}|k}}e^{-ikt}.\label{eq:flatspacemode}
\end{equation}
This function is related to $\xi$ by $N_{k}=\frac{m}{a}V=\frac{m}{a^{2}}\xi$.
The mode $N_{k}$ is defined in Minkowski spacetime, where $a\equiv1$ with
$t\equiv\tau,$ so we only need to follow a factor of $m$. Using the
asymptotic formula
\begin{equation}
\lim_{-c_{s}^{{(\pm1)}}k\tau\rightarrow\infty}H_{\nu}^{(1,2)}(-c_{s}^{{(\pm
1)}}k\tau)=\sqrt{\frac{2}{\pi}}\frac{1}{\sqrt{-c_{s}^{{(\pm1)}}k\tau}}e^{\mp
i(c_{s}^{{(\pm1)}}k\tau+\delta)},
\end{equation}
with $\delta=\frac{\pi}{2}(\nu+1/2)$, we find that in the sub-horizon limit,
\begin{equation}
\xi\rightarrow\frac{1}{\sqrt{2c_{s}^{{(\pm1)}}k}}\left[  \alpha_{k}
e^{-i(c_{s}^{{(\pm1)}}k\tau+\delta)}+\beta_{k}e^{+i(c_{s}^{{(\pm1)}}
k\tau+\delta)}\right]  .
\end{equation}
Matching to \cref{eq:flatspacemode}, and ignoring the unimportant phase
factors $e^{\pm i\delta}$, we see that we need
\begin{align}
\alpha_{k} &  =\frac{1}{4m}\sqrt{\frac{2}{|c_{1}|}},\\
\beta_{k} &  =0,
\end{align}
where we have (consistently) put in some factors of $c_{s}^{{(\pm1)}}$ which
do not appear in the flat spacetime calculation because it ignores gravity,
but would have appeared if we had included gravity.\footnote{To see this,
consider \cref{eq:xieom} in the case $a=1$, which is the spin-1 perturbation
equation in flat spacetime with gravitational perturbations turned on. Since
this requires $\phi=0$, the equation of motion (\ref{eq:xieom}) just becomes
$\xi^{\prime\prime}+c_{s}^{{(\pm1)}2}k^{2}\xi=0 $. This has the same solution
as we found in the case with gravity turned off in \cref{sec:flatspace}, but
with the sound speed modified, as expected.} Substituting in this value of
$\alpha_{k}$, we find the full solution for the spin-1 perturbation
\begin{equation}
\boxed{V^\pmvec = \frac{1}{a} \sqrt\frac{\pi}{2} \frac{1}{4m\sqrt{|c_1|}} \sqrt{-\tau} H_\nu^{(1)}(-c_s^\pmvec k\tau).}\label{eq:vksol}
\end{equation}
As a consistency check, if we turn off the scalar-aether coupling, we have
$\nu=1/2$, and (up to an irrelevant phase of $-\pi/2$) we recover equation
(91) in \cite{Lim:2004js}.

\subsection{Tachyonic Instability in the Vector Modes}

\label{sec:instability}

On super-horizon scales, the Hankel functions behave as
\begin{equation}
\lim_{c_{s}^{{(\pm1)}}k\tau\rightarrow0}H_{v}^{(1)}(-c_{s}^{{(\pm1)}}
k\tau)=\frac{i}{\pi}\Gamma(\nu)\left(  \frac{-c_{s}^{{(\pm1)}}k\tau}
{2}\right)  ^{-\nu}.
\end{equation}
Plugging this into \cref{eq:vksol}, we see that the large-scale vector perturbations to
the aether and metric depend on time as
\begin{equation}
V^{{(\pm1)}}\sim a^{-1}\sqrt{-\tau}(-\tau)^{-\nu}\sim(-\tau)^{\frac{3}{2}-\nu}.
\end{equation}

When the aether-scalar coupling (proportional to $\Lambda$) is small or absent, such that $-1/4<\Lambda<2$, the vector perturbations decay and are unobservable, as in pure \ae-theory \cite{Lim:2004js}. If $\Lambda$ is outside that range, then the coupling is large enough to change the nature of the vector perturbations. The coupling has two possible effects, depending on its sign. If $\nu$ is imaginary ($\Lambda < -1/4$), then the vector modes are both oscillatory and decaying.\footnote{Recall that, during inflation, $\tau$ runs from $-\infty$ to 0.} This corresponds to a large coupling which significantly damps the perturbations. On the other hand, the vector modes will experience runaway growth if $3/2-\nu$ is real and negative, or $\Lambda>2$. In this case the coupling is large, but with the opposite sign to the previous case, and this large coupling drives runaway production of aether modes. This is precisely the tachyonic instability we anticipated in \cref{sec:vecslowroll}, as it results from the aether perturbations acquiring an imaginary effective mass.

Since this growth is exponential (in cosmic time, or in number of $e$-foldings), it seems quite probable that this growing vector mode will overwhelm the slow-roll background solution and therefore lead to an instability. In this subsection we will calculate the growth of a single vector mode and compare it to the background evolution.

In order to maintain a homogeneous and isotropic background spacetime, the
time-space term in the stress-energy tensor must be zero at the level of the
background ($\bar T^{0}{}_i = 0$). The spin-1 perturbations do contribute to these terms in the
stress-energy tensor (\ref{eq:spin10i}) through terms proportional to
$V_{k}^{{(\pm1)}}Y_{i,k}^{{(\pm1)}}$. In particular, we will focus on the
scalar-aether coupling term
\begin{equation}
T^{0}{}_{i,k} = \ldots + m\bar{V}_{\theta\phi}\frac{\bar{\phi}^{\prime}}{a}
V_{k}^{{(\pm1)}}Y_{i,k}^{{(\pm1)}} + \ldots,
\end{equation}
which we will write as
\begin{equation}
T^{0}{}_{i,k}\supset m\bar{V}_{\theta\phi}\frac{\bar{\phi}^{\prime}}{a}
V_{k}^{{(\pm1)}}Y_{i,k}^{{(\pm1)}}.
\end{equation}

Our strategy will be to focus on a single mode, picking one of the larger modes available to us. Because $V_{k}$ grows with decreasing $k$, we choose a mode which crosses the sound horizon at some early conformal time $\tau_{i}$. Such a mode has wave number
\begin{equation}
k=\frac{1}{-c_{s}^{{(\pm1)}}\tau_{i}}.
\end{equation}
The perturbation $V_{k}^{{(\pm1)}}$ is given by \cref{eq:vksol}, which for a
super-horizon perturbation becomes
\begin{equation}
V_{k}^{{(\pm1)}}(N)=-\frac{i}{2\pi}\frac{H}{4m\sqrt{|c_{1}|}}\Gamma(\nu
)2^{\nu}\left(  -\tau_{i}\right)  ^{\frac{3}{2}}e^{\left(  \nu-\frac{3}
{2}\right)  N},
\end{equation}
where $N$ is the number of $e$-folds after the mode crossed the sound horizon.

The mode function $Y^{(\pm1)}_{i,k}$ is given by \cite{Lim:2004js}
\begin{equation}
Y^{(\pm1)}_{i,k}(\vec{x}) = \frac{1}{\sqrt{2}k}\left[ \left( \vec{k}\times
\vec{n}\right) _{i} \mp i \left( \vec{k}\times\vec{n}\right) _{i}\right]
e^{i\vec{k}\cdot\vec{x}},
\end{equation}
where $\vec{n}$ is a unit vector orthogonal to $\vec{k}$. We can always choose
three orthogonal coordinates such that $k^{i} = k\delta^{i}{}_{1}$ and
$n^{i}=\delta^{i}{}_{2}$, so the mode function is
\begin{equation}
Y^{(\pm1)}_{i,k}(\vec{x}) = \frac{1\mp i}{\sqrt{2}}e^{i\vec{k}\cdot\vec{x}
}\delta^{3}{}_{i}.
\end{equation}
This oscillates throughout space; we will choose $\vec{x}$ such that
$\operatorname{Re}\left[ (i\pm1)e^{i\vec{k}\cdot\vec{x}}\right] $ has its
maximum value of 1. (The other terms in $V_{k}^{(\pm1)} Y^{(\pm1)}_{i,k}$ are
all manifestly real.)

Therefore, this particular mode has a contribution to the $0-i$ component of
the stress-energy tensor which includes a term
\begin{equation}
T^{0}{}_{i,k}\supset-m\bar{V}_{\theta\phi}\frac{\bar{\phi}^{\prime}}{a}
\frac{1}{2\pi}\frac{H}{4m\sqrt{|c_{1}|}}\Gamma(\nu)2^{\nu}\left(  -\tau
_{i}\right)  ^{\frac{3}{2}}e^{\left(  \nu-\frac{3}{2}\right)  N}\frac{1}
{\sqrt{2}}\delta^{3}{}_{i}.
\end{equation}
Using the slow-roll equation for $\bar{\phi}$, $3\sh\bar\phi' \approx -a^2 \bar V_\phi$, we can write this as
\begin{equation}
T^{0}{}_{i,k}\supset\frac{\bar{V}_{\phi}\bar{V}_{\theta\phi}}{24\pi
\sqrt{|c_{1}|}}\Gamma(\nu)2^{\nu-\frac{1}{2}}\left(  -\tau_{i}\right)
^{\frac{3}{2}}e^{\left(  \nu-\frac{3}{2}\right)  N}\delta^{3}{}_{i}.
\end{equation}
Comparing this to the background $0-0$ component of $T^{\mu}{}_{\nu}$,
$\bar{T}^{0}{}_{0}=\bar{\rho}=3H^{2}/8\pi\tilde G_{c}$, we find
\begin{equation}
T^{0}{}_{i,k}/\bar{T}^{0}{}_{0}\supset\frac{\tilde G_{c}\bar{V}_{\phi}\bar
{V}_{\theta\phi}}{9H^{2}\sqrt{|c_{1}|}}\Gamma(\nu)2^{\nu-\frac{1}{2}}\left(
-\tau_{i}\right)  ^{\frac{3}{2}}e^{\left(  \nu-\frac{3}{2}\right)  N}
\delta^{3}{}_{i}.\label{eq:t0icomp}
\end{equation}
Using the slow-roll Friedmann equation we could re-write this purely in terms
of the potential as
\begin{equation}
T^{0}{}_{i,k}/\bar{T}^{0}{}_{0}\supset\frac{1}{24\pi\sqrt{|c_{1}|}}\frac
{\bar{V}_{\phi}\bar{V}_{\theta\phi}}{\bar{V}-\bar{\theta}\bar{V}_{\theta}
}\Gamma(\nu)2^{\nu-\frac{1}{2}}\left(  -\tau_{i}\right)  ^{\frac{3}{2}
}e^{\left(  \nu-\frac{3}{2}\right)  N}\delta^{3}{}_{i}.\label{eq:t0icomppot}
\end{equation}

The key feature here is the exponential dependence on $N$ for $\nu>3/2$ (the
condition we found above for $V_{k}^{{(\pm1)}}$ to grow exponentially in cosmic time). While
the derivatives of the potential in the numerator of \cref{eq:t0icomppot}
should be a few orders of magnitude smaller than the potential in the
denominator, as a consequence of slow roll, this is likely to be dwarfed by
the exponential dependence on the number of $e$-folds, which even for the bare
minimum length of inflation, $N\sim50$--$60$, will be very large. Moreover, as
we will see in \cref{sec:dj}, $\nu$ can in principle be larger than $3/2$ even
by several orders of magnitude, hence the other terms with exponential
dependence on $\nu$, as well as the gamma function, can be quite large as well.

Therefore, when $\nu>3/2$ the vector modes will generically drive the
off-diagonal term in the stress-energy tensor far above the background
density. This does not necessarily mean that isotropy is violated. As discussed in \cref{sec:spin0perts}, the same
physical process that drives $V_{k}^{{(\pm1)}}$ will similarly pump
energy into the spin-0 piece, $V_{k}^{(0)}$, which affects the
perturbations to $\bar{T}^{0}{}_{0}$ as well as $\bar{T}^{0}{}_{i}$. Consequently, background
homogeneity and isotropy could still hold, but the slow-roll solution to the
background Friedmann equations which we perturbed would be invalid. Either way
our inflationary background becomes dominated by the perturbations.

Note that this calculation was done for a single mode, albeit one of the
largest ones available because $V_{k}$ grows for smaller $k$. Integrating over
all modes produced during inflation would of course exacerbate the
instability.

This instability is explored in greater quantitative detail in
\cref{sec:dj}, where we examine a specific potential for which we can
elucidate the constraints on $\bar{V}_{\phi}$ and $\bar{V}_{\theta\phi}$.

\subsection{What Values Do We Expect for \texorpdfstring{$\Lambda$}{Lambda}?}

\label{sec:lambdavalues}

$V^{{(\pm1)}}$ has an effective mass-squared (\ref{eq:meffsq}) which depends on both the theory's free parameters and derivatives of the scalar potential, and can be of either sign. When it is negative, the aether modes are tachyonic and $V^\pmvec$ contains an exponentially growing mode. This occurs when the parameter $\Lambda$, defined in \cref{eq:lambdadef}, satisfies $\Lambda>2$. To lowest order in
slow-roll, $\Lambda$ is written in terms of several free parameters:
$c_{1},m,H,$ and the potential derivatives $\bar{V}_{\theta\phi}$ and $\bar
{V}_{\phi}$:
\begin{equation}
\Lambda \equiv -\frac{\bar{V}_{\phi}\bar{V}_{\theta\phi}}{6mc_{1}H^{3}
}+\mathcal{O}(\varepsilon).
\end{equation}
Hence, $\Lambda$ can span a fairly large range of orders of magnitude. However, there are several existing constraints on these parameters, most of
which constrain several of them in terms of each other.

There are two things we can do to clarify this expression for $\Lambda$. We generally expect that for a slow-roll phase, $\ddot{\phi}\ll3H\dot{\phi}$
and $\frac{1}{2}\dot{\phi}^{2}\ll M_{\mathrm{Pl}}^{2}H^{2}$, where the Planck
mass is given as usual by
\begin{equation}
M_{\mathrm{Pl}}^{-2}=8\pi G=8\pi G_{c}\cdot\mathcal{O}(1).
\end{equation}
We will rewrite the second inequality in terms of a slow-roll
parameter, $\zeta$, as
\begin{align}
\frac{1}{2}\dot{\phi}^{2} &  =M_{\mathrm{Pl}}^{2}H^{2}\zeta,\\
\zeta &  \ll1.
\end{align}
Using the slow-roll Friedmann and Klein-Gordon equations, we
can then re-write $\Lambda$ as
\begin{equation}
\Lambda = \operatorname{sgn}(\dot{\phi})\frac{\zeta^{1/2}}{\sqrt{2}c_{1}
}\left(  \frac{m}{M_{\mathrm{Pl}}}\right)  ^{-1}\frac{\bar{V}_{\theta\phi}}
{H}+\mathcal{O}(\varepsilon).
\end{equation}

Next, we can re-define the coupling $\bar V_{\theta\phi}$ using the flat spacetime stability constraint
(\ref{eq:spin0constraint}) that we derived in \cref{sec:flatspace}. If we
define a normalized coupling $\sigma$ by
\begin{equation}
V_{\theta\phi}^{2}(0,0)\equiv2c_{123}M_{0}^{2}\sigma,
\end{equation}
where $M_{0}^{2}=V_{\phi\phi}(0,0)$ is the effective mass-squared of the
scalar around a Minkowski background, then the stability constraint is simply
\begin{equation}
\sigma\leq1.
\end{equation}
Therefore, we have
\begin{equation}
\Lambda = \operatorname{sgn}(\dot{\phi})\zeta^{1/2}\sigma^{1/2}\frac
{c_{s}^{(0)}}{\sqrt{c_{1}}}\frac{\bar{V}_{\theta\phi}}{V_{\theta\phi}
(0,0)}\frac{M_{0}}{H}\left(  \frac{m}{M_{\mathrm{Pl}}}\right)  ^{-1}
+\mathcal{O}(\varepsilon).\label{eq:lambda}
\end{equation}

The instability occurs when $\Lambda>2$. Let us first examine
\cref{eq:lambda} to see if it can be positive. Most of the terms are
manifestly positive. Positivity of the Hamiltonian for spin-1 perturbations in
flat spacetime requires $c_{1}\geq0$ \cite{Lim:2004js}.\footnote{This was derived in pure \ae-theory. However, recall from
\cref{sec:flatspace} that the spin-1 modes in flat spacetime are unaffected by
the scalar-aether coupling, as $\theta$ is a divergence of the vector
perturbations and so only contains spin-0 perturbations. Hence, $c_{1}$ needs
still be positive.} Tachyonic stability of the scalar requires $M_{0}$ to be
real and positive. The timelike constraint on the aether requires $m$ be
positive as well. Putting this all together, we find
\begin{equation}
\Lambda = \operatorname{sgn}(\dot{\phi})\frac{\bar{V}_{\theta\phi}}
{V_{\theta\phi}(0,0)}\underbrace{\zeta^{1/2}
\vphantom{\frac{c_s^{(0)}}{\sqrt{c_1}}}}_{>0}\underbrace{\sigma^{1/2}
\vphantom{\frac{c_s^{(0)}}{\sqrt{c_1}}}}_{>0}\underbrace{\frac{c_{s}^{(0)}
}{\sqrt{c_{1}}}}_{>0}\underbrace{\frac{M_{0}}{H}
\vphantom{\frac{c_s^{(0)}}{\sqrt{c_1}}}}_{>0}\underbrace{\left(  \frac
{m}{M_{\mathrm{Pl}}}\right)  ^{-1}\vphantom{\frac{c_s^{(0)}}{\sqrt{c_1}}}}
_{>0}+\mathcal{O}(\varepsilon),
\end{equation}
implying that in order for $\Lambda$ to be positive, $\dot{\phi}$ and the
coupling $\bar{V}_{\theta\phi}$ need to have the same sign. This is not
difficult to achieve in practice; in the quadratic potential of
\cite{Donnelly:2010cr} (see \cref{sec:dj} for more discussion), it amounts to
requiring that $\phi$ and $\bar{V}_{\theta\phi}$ have opposite signs, which is
true for a large space of initial conditions leading to inflating trajectories.

Next we need to see under which conditions $|\Lambda|$ can be $\mathcal{O}(1)$
or greater. We have assumed that the scalar slow-roll parameter $\zeta$ is
small. In particular, in the absence of the aether, $\zeta$ is equal to
$\varepsilon$, which observations constrain to be $\sim\mathcal{O}(10^{-2})$ \cite{Ade:2013zuv}. It
therefore seems sensible that $\zeta^{1/2}$ should be small but not terribly
small, perhaps $\sim\mathcal{O}(10^{-1})$ or so.

Similarly, the scalar-aether coupling $\sigma$ is constrained by flat
spacetime stability of the spin-0 modes to be strictly less than 1. However,
we do not want to consider couplings so small as to be uninteresting, so we
may choose the coupling to be as close to $\sigma=1$ as is allowed. Therefore,
$\sigma^{1/2}$ ought to be smaller, but need not be too much smaller, than 1.

Written in the form of \cref{eq:lambda}, the value of $\Lambda$ is sensitive
to how $V_{\theta\phi}$ and $V_{\phi\phi}$ differ between a quasi-de Sitter
inflationary background and a Minkowski background. In the quadratic potential
$V(\theta,\phi)=\frac{1}{2}M^{2}\phi^{2}+\mu\theta\phi$ which we discuss in
\cref{sec:dj}, both of these are constant, although one could presumably
construct inflationary potentials for which this is not true. The effective
mass of the scalar during inflation, $M=\bar{V}_{\phi\phi}^{1/2}$, should be
less than the Hubble rate in order to produce perturbations. Putting all this
together, we are left with
\begin{equation}
\Lambda = \operatorname{sgn}(\dot{\phi})\underbrace{\zeta^{1/2}
\vphantom{\frac{c_s^{(0)}}{\sqrt{c_1}}}}_{<1}\underbrace{\sigma^{1/2}
\vphantom{\frac{c_s^{(0)}}{\sqrt{c_1}}}}_{\leq1}\underbrace{\frac{c_{s}^{(0)}
}{\sqrt{c_{1}}}}_{\mathcal{O}(1)}\underbrace{\frac{\bar{V}_{\theta\phi}
}{V_{\theta\phi}(0,0)}\frac{M_{0}}{M}}_{\mathcal{O}(1)?}\underbrace{\frac
{M}{H}\vphantom{\frac{c_s^{(0)}}{\sqrt{c_1}}}}_{\ll1}\underbrace{\left(
\frac{m}{M_{\mathrm{Pl}}}\right)  ^{-1}
\vphantom{\frac{c_s^{(0)}}{\sqrt{c_1}}}}_{\gg1?}+\mathcal{O}(\varepsilon). \label{eq:lambdaoom}
\end{equation}
We can see that in order for $\Lambda$ to be larger than 2, the aether VEV,
$m$, needs to be at least a few orders of magnitude smaller than the Planck
scale. $m$ is effectively the Lorentz symmetry-breaking mass scale. It can
therefore be quite a bit smaller than the Planck mass, although if it were below the scale of collider experiments, any couplings to matter could displace the aether from its VEV and Lorentz violating effects could be visible.

There are several experimental and observational results suggesting that $m/M_\mathrm{Pl}$ should be quite small. Here we briefly discuss three strong constraints, arising from big bang nucleosynthesis, solar system tests, and the absence of gravitational \v{C}erenkov radiation, as well as a possible caveat.

As mentioned in \cref{sec:eom}, the gravitational constant appearing in the Friedmann equations, $\gc$, and the gravitational constant appearing in the Newtonian limit, $G_{N}$, are both displaced from the ``bare" gravitational constant, $G$, by a factor that is, schematically, $1+c_i(m/M_\mathrm{Pl})^2$. The primordial abundances of light elements such as helium and deuterium probe the cosmic expansion rate during big bang nucleosynthesis, which depends on $\gc$ through the Friedmann equations. Therefore, by comparing this to $G_N$ measured on Earth and in the solar system, $c_im^2$ can be constrained. Assuming the $c_i$ are $\mathcal{O}(1)$,\footnote{As the $c_i$ are dimensionless parameters, this is perfectly reasonable. Note that even if $m$ were order $M_\mathrm{Pl}$ or larger and the constraints discussed in here are actually constraints on the smallness of the $c_i$, $\Lambda$ still depends on these parameters as $c_1^{-1/2}$.} the BBN constraint implies $m/M_\mathrm{Pl}\lesssim10^{-1}$ \cite{Carroll:2004ai}.

Slightly better constraints on $\gc/G_N$ come from the cosmic microwave background (CMB) \cite{Robbers:2007ca,Bean:2010zq}. The tightest bound, $|G_N/\gc-1|<0.018$ at $95\%$ confidence level, was computed using CMB data (WMAP7 and SPT) and the galaxy power spectrum (WiggleZ) in a theory closely related to the one described in this paper, and should hold generally for \ae-theory at the order-of-magnitude level \cite{Audren:2013dwa}. These constrain $m/M_\mathrm{Pl}$ to be no greater than a few percent.

There are yet stronger bounds on $m/\Mp$ through constraints on the preferred-frame parameters, $\alpha_{1,2}$, in the parameterized post-Newtonian (PPN) formalism. These coefficients scale, to leading order, as $c_i(m/M_\mathrm{Pl})^2$ \cite{Foster:2005dk,Jacobson:2008aj}. The observational bounds $\alpha_1\lesssim10^{-4}$ and $\alpha_2\lesssim4\times10^{-7}$ therefore imply $m/M_\mathrm{Pl} \lesssim 6\times10^{-4}$. Recent pulsar constraints on $\alpha_{1,2}$ are even stronger than this \cite{Shao:2013wga}, although they are derived in the strong-field regime and thus might not be directly applicable to the weak-field \ae-theory results. Similarly, recent binary pulsar constraints on Lorentz violation \cite{Yagi:2013qpa} constrain $m/\Mp\lesssim10^{-1}$, assuming $c_i \sim\mathcal{O}(1)$.

The strongest constraints come from the absence of \textquotedblleft
gravitational \v{C}erenkov radiation." Because the aether changes the permeability of the vacuum, coupled aether-graviton modes may travel subluminally, despite being nominally massless. Consequently, high-energy particles moving at greater speeds can emit these massless particles, in
analogy to the usual \v{C}erenkov radiation. This emission causes high-energy
particles to lose energy, and at an increasing rate for higher energy particles.
Among the highest energy particles known are cosmic rays, which travel
astronomical distances and hence could degrade drastically due to
such gravitational \v{C}erenkov effects. Such a degradation has, however, not been observed; this generically constrains
$m/M_{\mathrm{Pl}}<3\times10^{-8}$ \cite{Elliott:2005va}.

We should note that these constraints can be side-stepped if certain convenient exact relationships hold among the $c_{i}$, although crucially they cannot all be avoided in this way simultaneously without allowing for superluminal propagation of the aether modes \cite{Jacobson:2008aj}. The PPN parameters $\alpha_{1,2}$ are identically zero when $c_3=0$ and $2c_1=-3c_2$. The BBN constraint is automatically satisfied by requiring $2c_1 + 3c_2 + c_3$ to vanish, as this sets $\gc=G_N$. \cite{Carroll:2004ai}. Note that the PPN cancellations imply the BBN cancellation, though the reverse is not necessarily true.\footnote{The conditions for PPN and BBN to cancel can be relaxed by including a $c_4$ term which describes a quartic aether self-interaction. We have ignored such a term in order to simplify the theory, although like the other three terms, it is permitted when that the aether equations of motion are demanded to be second order in derivatives. When $c_4 \neq0$, the vanishing of $\alpha_{1,2}$ continues to imply that the BBN constraints are satisfied.} The \v{C}erenkov constraints vanish if all five propagating gravitational (metric and aether) degrees of freedom propagate exactly luminally. This happens when $c_3=-c_1$ and $c_2=c_1/(1-2c_1)$ \cite{Elliott:2005va}. Note that while $\alpha_2=0$ in this parameter subspace, $\alpha_1 = -8c_1(m/M_\mathrm{Pl})^2$, which would place a constraint on $m/M_\mathrm{Pl}$ of order $10^{-2}$. It is worth mentioning that the \v{C}erenkov constraints on $m$ will also be avoided if the mode speeds for some of the aether-metric modes are superluminal. This includes a two-dimensional parameter subspace in which the PPN and BBN constraints are automatically satisfied \cite{Jacobson:2008aj}. Whether superluminal propagation is acceptable in \ae-theory is somewhat controversial. It is a metric theory of gravity, so superluminality should imply violations of causality, including propagation of energy around closed timelike curves \cite{Elliott:2005va,Lim:2004js}. However, this may be seen as an \textit{a posteriori} demand, and some authors (e.g., \cite{Jacobson:2008aj}) do not require it.

It is unclear what fundamental physical principle, if any, would cause the $c_i$ to cancel in any of the aforementioned ways. Hence it seems to be a fairly general result that $m$ must be many orders of magnitude below the Planck scale. If $m/M_\mathrm{Pl}$ is small enough compared to $M/H$ and the other small parameters appearing in \cref{eq:lambdaoom}, $\Lambda$ can easily be above 2 and the aether-inflaton coupling runs a serious danger of causing an instability. For a given $m/\Mp$, this places a constraint on the size of the coupling, $\bar V_{\theta\phi}$. We will discuss this constraint more quantitatively in \cref{sec:dj} for a specific choice of the potential.

\section{Spin-0 Cosmological Perturbations: Instability and Observability}
\label{sec:spin0perts}

We now consider the spin-0 perturbations. For readers who want to skip the
calculational details, we first summarize this section. The spin-0 equations
are complicated by the addition of $\delta \phi $ modes which add a new degree of freedom. In order to tackle
these equations, we use the smallness of $m/M_{\mathrm{Pl}}$, discussed in 
\cref{sec:lambdavalues}, to solve the perturbations order-by-order, along
the lines of the approach in \cite{Blas:2011en}. At lowest order in $m/M_{
\mathrm{Pl}}$, the perturbations $\Phi $ and $\delta \phi $ have the same
solutions as in the standard slow-roll inflation in general relativity.
These can be substituted into the $\xi $ equation of motion to solve for $
\xi $ at lowest order, which we then substitute back into the $\Phi $ and $
\delta \phi $ equations at $\mathcal{O}(m/M_{\mathrm{Pl}})$.

The instability found in the spin-1 perturbations reappears, and occurs in
essentially the same region of parameter space. We then assume that the
parameters are such that this instability is absent, in which case $\xi$ is
roughly constant. We solve for the metric perturbation $\Phi$ and find that
neither its amplitude nor scale-dependence are significantly changed from
the standard slow-roll case. In particular, we calculate two key
inflationary observables: the scale-dependence of the $\Phi$ power spectrum, 
$n_s$, and the tensor-to-scalar ratio, $r$.

Surprisingly, the first corrections due to the aether-scalar coupling enter
at $\mathcal{O}(m/M_\mathrm{Pl})^2$. Up to first order in $m/M_\mathrm{Pl}$,
the aether-scalar coupling has no effect on cosmic perturbations on
super-horizon scales, assuming that $m/M_\mathrm{Pl}$ is small compared to
unity and that the perturbations are produced during a slow-roll quasi-de
Sitter phase. A corollary of this is that super-horizon isocurvature modes, a
generic feature of coupled theories, are not produced by the aether-scalar
coupling up to $\mathcal{O}(m/M_\mathrm{Pl})^2$. Because of the smallness of 
$m/M_\mathrm{Pl}$, any deviations to $n_s$ and $r$ caused by the
aether-scalar coupling are unobservable to the present and near-future
generation of CMB experiments.

Since the pure \ae -theory terms in the perturbed Einstein equations carry
two powers of $u^{\mu }$ (which is proportional to $m$) and so only begin to
contribute at $\mathcal{O}(m/M_{\mathrm{Pl}})^{2}$, we will not recover the
cosmological perturbation results of pure \ae -theory by taking any limits,
as we only work in this section to $\mathcal{O}(m/M_{\mathrm{Pl}})$. The
effects of \ae -theory on the spin-0 perturbations are mild, amounting
essentially to a rescaling of the power spectrum amplitude that is $\mathcal{
O}(m/M_{\mathrm{Pl}})^{2}$ and is degenerate with $\varepsilon $ \cite
{Lim:2004js}.

\subsection{The Spin-0 Equations of Motion}

In order to eliminate non-physical degrees of freedom, we need to specify a
choice of coordinate system with no remaining gauge freedom. We choose to
work in Newtonian gauge, where $B^{(0)}=H_{T}^{(0)}=0$. The equations of
motion are relatively simple in this gauge, and the perturbation $\Phi $ has
a simple interpretation as the relativistic generalization of the Newtonian
gravitational potential \cite{Mukhanov:1990me}. Hereafter we will drop the
spin-0 superscripts.

The $0-0$, $0-i$, and $i-i$ Einstein equations, respectively, are 
\begin{align}
4\pi\tilde G_c\left(-\bar\phi'\delta\phi'-a^2\bar V_\phi\delta\phi\right) &= \left(3\sh^2 - A\right)\Phi - 3\mathcal{H}\Psi' - \frac{\tilde G_c}{G}k^2\Psi - 8\pi\tilde G_cc_1m^2k^2\Phi \nonumber \label{eq:spin000} \\
&\hphantom{{}=} + 8\pi\tilde G_cc_1m^2k(V'+\mathcal{H}V) - 8\pi\tilde G_c\tilde\alpha\mathcal{H}kV \nonumber \\
&\hphantom{{}=} + 4\pi\tilde G_cma\bar V_{\theta\phi}(\bar\phi'\Phi-3\mathcal{H}\delta\phi) \\
\frac{1}{8\pi G}(k\mathcal{H}\Phi-k\Psi') &= \frac{k}{2}\bar\phi'\delta\phi - \tilde\alpha AV +c_1m^2a^{-1}(ak\Phi)' \nonumber \label{eq:spin00i} \\
&\hphantom{{}=} - c_1m^2\frac{\xi''}{a} + \frac{1}{2}ma\bar V_{\theta\phi}\bar\phi'V \\
4\pi\tilde G_c\left(\bar\phi'\delta\phi'-a^2\bar V_\phi\delta\phi\right) &= \left(3\sh^2 - A\right)\Phi + \mathcal{H}\Phi' - 2\mathcal{H}\Psi' - \Psi'' - \frac{8\pi\tilde G_cm^2}{\gamma}\tilde c_{123}k^2(\Phi+\Psi) \label{eq:spin0ij} \nonumber \\
&\hphantom{{}=} + 4\pi\tilde G_c\frac{3m^3}{a}A\bar V_{\theta\theta\theta}\left(3\Psi^\prime-3\mathcal{H}\Phi+kV\right). \nonumber \\
&\hphantom{{}=} - 4\pi\tilde G_cma\left[\bar V_{\theta\phi}(3\mathcal{H}\delta\phi+\delta\phi') + \bar V_{\theta\phi\phi}\bar\phi'\delta\phi\right] \nonumber \\
&\hphantom{{}=} + 4\pi\tilde G_cm^2\bar V_{\theta\theta\phi}\left[3A\delta\phi-\bar\phi'(3\Psi'-3\mathcal{H}\Phi+kV)\right].
\end{align}
The off-diagonal $i-j$ Einstein equation, unmodified by the coupling between
the aether and scalar, gives a constraint, 
\begin{equation}
k^{2}(\Phi +\Psi )=\gamma a^{-2}(a^{2}kV)^{\prime },  \label{eq:offdiagspin0}
\end{equation}
where $\gamma \equiv 16\pi Gm^{2}c_{13}$ was defined in \cref{sec:spin1perts}. We may eliminate $\Psi $ and its derivatives by the constraint (\ref{eq:offdiagspin0}) and its conformal time derivatives, 
\begin{align}
\Psi' &= \gamma(ak)^{-1}\left(\xi''-A\xi\right)-\Phi', \label{eq:offdiagspin0der1} \\
\Psi'' &= \gamma(ak)^{-1}\left(\xi'''-\mathcal{H}\xi''-A\xi'+A\xi\left(\mathcal{H}-\frac{A'}{A}\right)\right)-\Phi'', \label{eq:offdiagspin0der2}
\end{align}
where, as for the spin-1 perturbations, we have defined $\xi \equiv aV$ and $
A={\mathcal{H}}^{2}-{\mathcal{H}}^{\prime }$. Note the presence of third
derivatives of $\xi $ in the expression for $\Psi ^{\prime \prime }$, which
could severely complicate the Einstein equations at $\mathcal{O}(m/M_{
\mathrm{Pl}})^{2}$.

Finally, the $\nu =i$ aether equation of motion is, using 
\cref{eq:offdiagspin0,eq:offdiagspin0der1,eq:offdiagspin0der2}, 
\begin{equation}
\xi ^{\prime \prime }+\frac{\tilde{c}_{123}m^{2}}{c_{1}m^{2}+\tilde{\alpha}
\gamma }k^{2}\xi +\left( \frac{\tilde{\alpha}(1-\gamma )A-\frac{1}{2}ma\bar{V
}_{\theta \phi }\bar{\phi}^{\prime }}{c_{1}m^{2}+\tilde{\alpha}\gamma }
\right) \xi =\frac{c_{1}m^{2}+\tilde{\alpha}}{c_{1}m^{2}+\tilde{\alpha}
\gamma }k(a\Phi )^{\prime }-\frac{1}{2}\frac{ma^{2}\bar{V}_{\theta \phi }}{
c_{1}m^{2}+\tilde{\alpha}\gamma }k\delta \phi  \label{eq:xispin0eom}
\end{equation}
where, as before, tildes indicate the usual \ae -theory constants modified
by appropriate factors of $\frac{1}{2}\bar{V}_{\theta \theta }$.

We can perform a consistency check by observing that these reduce to $\delta
T^{\mu }{}_{\nu }$ for a single scalar field in general relativity \cite
{Mukhanov:1990me} when the aether is turned off (in the limit $m\rightarrow
0 $), as well as $\delta T^{\mu }{}_{\nu }$ and the $\xi $-equation of
motion in \ae -theory \cite{Lim:2004js} in the limit $V(\theta ,\phi
)\rightarrow V(\phi )$.

\subsection{The Instability Returns}

\label{sec:spin0instability}

To lowest order in $m/M_{\mathrm{Pl}}$, the constraint equation (\ref
{eq:offdiagspin0}) tells us simply that the anisotropic stress vanishes: $
\Psi =-\Phi $. Taking this into account, the $0-i$ Einstein equation at
lowest order in $m/M_{\mathrm{Pl}}$ is 
\begin{equation}
\left( a\Phi \right) ^{\prime }=4\pi Ga\bar{\phi}^{\prime }\delta \phi .
\label{eq:0iozero}
\end{equation}
The $\nu =i$ aether equation of motion (\ref{eq:xispin0eom}) is, dropping
terms of $\mathcal{O}(m^{2}/M_{\mathrm{Pl}}^{2})$, 
\begin{equation}
\xi ^{\prime \prime }+c_{s}^{(0)2}k^{2}\xi -\frac{a\bar{V}_{\theta \phi }
\bar{\phi}^{\prime }}{2mc_{1}}\xi =\left( 1+\frac{\alpha }{c_{1}m^{2}}
\right) k(a\Phi )^{\prime }-\frac{a^{2}\bar{V}_{\theta \phi }k}{2mc_{1}}
\delta \phi
\end{equation}
where 
\begin{equation}
c_{s}^{(0)2}=\frac{{c}_{123}m^{2}}{c_{1}m^{2}+\alpha \gamma }=\frac{c_{123}}{
c_{1}}\left(1+\mathcal{O}\left( \frac{m}{M_{\mathrm{Pl}}}\right) ^{2}\right)
\end{equation}
is the same spin-0 sound speed as in flat space (cf. \cref{sec:flatspace})
to first order in $m/M_{\mathrm{Pl}}$. In de Sitter spacetime this becomes,
using \cref{eq:0iozero} to replace $(a\Phi )^{\prime }$ with $\delta \phi $, 
\begin{equation}
\xi ^{\prime \prime }+c_{s}^{(0)2}k^{2}\xi -\frac{\bar{V}_{\theta \phi }\dot{
\bar{\phi}}}{2mc_{1}H^{2}}\frac{\xi }{\tau ^{2}}=\frac{k}{H^{2}}\left[
\left( 1+\frac{\alpha }{c_{1}m^{2}}\right) 4\pi G\dot{\bar{\phi}}-\frac{1}{2}
\frac{\bar{V}_{\theta \phi }}{mc_{1}}\right] \frac{\delta \phi }{\tau ^{2}},
\label{eq:spin0xieom}
\end{equation}
to lowest order in the slow-roll parameters and $m/M_{\mathrm{Pl}}$.

Combined with the perturbed Klein-Gordon equation, $\xi $ and $\delta \phi $
obey coupled oscillator equations. However, to zeroth order in $m/M_{\mathrm{
Pl}},$ the scalar field is unaffected by the aether perturbations,\footnote{The aether coupling will still enter the perturbed Klein-Gordon equation at
this order through the potential terms.} so on super-horizon scales $\delta
\phi $ is constant up to slow-roll corrections, resulting in the standard
nearly scale-invariant power spectrum. This is consistent with the flat space case discussed in \cref{sec:flatspace}, where it was found that the coupling to the aether does not destabilize the scalar modes. Therefore, on super-horizon scales, $
c_{s}^{(0)}k\tau \ll 1$, \cref{eq:spin0xieom} is solved by 
\begin{equation}
\xi =C_{+}\tau ^{n_{+}}+C_{-}\tau ^{n_{-}}+k\delta \phi \left[ \dot{\bar{\phi
}}^{-1}-\left( 1+\frac{\alpha }{c_{1}m^{2}}\right) \frac{m}{M_{\mathrm{Pl}}}
\frac{c_{1}}{\bar{V}_{\theta \phi }M_{\mathrm{Pl}}}\right] , \label{eq:spin0xisol}
\end{equation}
where $C_{\pm }$ are arbitrary constants, and 
\begin{equation}
n_{\pm }=\frac{1}{2}\pm \sqrt{\frac{1}{4}+\frac{\bar{V}_{\theta \phi }\dot{
\bar{\phi}}}{2mc_{1}H^{2}}}.
\end{equation}
As with the spin-1 perturbations, the spin-0 piece of $V=\xi /a$ can either
grow or decay exponentially (in cosmic time). In this case it will grow if 
\begin{equation}
\frac{\bar{V}_{\theta \phi }\dot{\bar{\phi}}}{2mc_{1}H^{2}}>2.
\end{equation}
This is exactly the same as the condition, $\Lambda >2$, for the spin-1
modes to be unstable. The real condition for instability may be slightly
different, as $\Lambda >2$ could violate our assumption that $m/M_{\mathrm{Pl
}}$ is small; however, the additional $\mathcal{O}(m/M_{\mathrm{Pl}})^{2}$
terms would only change some small multiplicative factors, and not by orders
of magnitude.

As in the spin-1 case, we can most easily see the effect of unstable aether
modes on the metric perturbations through the off-diagonal $i-j$ Einstein
equation (\ref{eq:offdiagspin0}). If $V$ blows up exponentially then so will 
$\Phi +\Psi $, and the metric perturbations will overwhelm the
Friedmann-Robertson-Walker background.

\subsection{The Small Coupling Limit}

Henceforth, we will assume that the aether perturbations are stable, so that
\begin{equation}
\Lambda \equiv \frac{\bar{V}_{\theta \phi }\dot{\bar{\phi}}}{2mc_{1}H^{2}}<2.
\end{equation}
This can be further split into two dominant cases, $|\Lambda|\ll1$ and $\Lambda<-1/4$. There are regions in parameter space which are not covered by these cases, such as $\Lambda \sim 1$, but these are likely to be highly fine-tuned as many of the parameters which enter $\Lambda$ have no relationship to each other \textit{a priori}. Consequently we should consider various values of $\Lambda$ on an order-of-magnitude basis.

$\Lambda \ll 1$ corresponds to the limit where the coupling $|\bar V_{\theta\phi}|$ is small compared to the mass scale $c_1mH^2/\dot{\bar\phi}$. Assuming that the background relations for the slow-roll parameters hold as in GR (which we will explore more rigorously in \cref{sec:dj} for a particular potential), then we have $\varepsilon = 4\pi G\dot{\bar{\phi}}^{2}/H^{2}$ up to $\mathcal{O}(m/\Mp)$, and this limit can be written as
\begin{equation}
\frac{|\bar V_{\theta\phi}|}{H} \ll \frac{c_1}{\sqrt\varepsilon}\frac{m}{\Mp}.
\end{equation}
In this limit, the term $C_- \tau^{n_-}$ is constant up to slow-roll corrections, as is the term proportional to $\delta\phi$. 

This case should be qualitatively similar to \ae-theory as it makes the aether-scalar coupling very small. However, we might be worried by the appearance of a $\bar V_{\theta\phi}^{-1}$ in \cref{eq:spin0xisol}. The limit $\bar V_{\theta\phi} \to 0$ does smoothly go to \ae-theory. The aether perturbation $\xi$ only appears, to $\mathcal{O}(m/\Mp)$, in the $0-i$ Einstein equation,
\begin{equation}
\Phi' + \sh\Phi = 4\pi G \bar\phi' \left(\delta\phi + \frac{m\bar V_{\theta\phi}}{k}\xi\right).
\end{equation}
The $\bar V_{\theta\phi}$ in the $\mathcal{O}(m/\Mp)$ term will cancel out the problematic $\bar V_{\theta\phi}^{-1}$ in the solution for $\xi$. Taking $\Lambda \to 0$ and substituting in the solution (\ref{eq:spin0xisol}), this becomes
\begin{equation}
\Phi' + \sh\Phi \approx 4\pi G\bar\phi'\delta\phi\left[1 - \frac{m^2}{\Mp^2}c_1\left(1+\frac{\alpha}{c_1m^2}\right)\right].
\end{equation}
The corrections enter at $\mathcal{O}(m/\Mp)^2$ and are negligible for the purposes of this analysis. Therefore the limit $|\Lambda| \ll 1$ should only differ from \ae-theory at $\mathcal{O}(m/\Mp)^2 \lesssim 10^{-15}$.

It is worth mentioning that for small but finite $\Lambda$ there will be new effects on extremely large scales, $k \lesssim \bar V_{\theta\phi}$. These may or may not be observable, depending on the scales covered during inflation.

\subsection{The Large Coupling Limit: The \texorpdfstring{$\Phi$}{Phi} Evolution Equation}

One interesting case is left: a large coupling with opposite sign to $\dot{\bar\phi}$, or $\Lambda < 1/4$. We will consider this for the rest of this section. However, we should mention that the sign of $\dot{\bar\phi}$ depends on initial conditions, and if this sign condition were not satisfied, then (as discussed in \cref{sec:spin0instability}) the aether-scalar coupling would drive a severe tachyonic instability. Hence such a large coupling may not be an ideal part of a healthy inflationary theory.

In this large coupling case, both of the $\tau ^{\pm }$ terms are decaying and we will take 
\begin{align}
\xi &= \frac{k\delta \phi }{\dot{\bar{\phi}}} \left(1+\mathcal{O}\left( \frac{m}{M_{
\mathrm{Pl}}}\right)\right) \label{eq:xisolspin0}\\
&\approx \frac{\sqrt{4\pi G}}{H\varepsilon ^{1/2}}k\delta \phi .
\end{align}
\Cref{eq:xisolspin0} was derived for super-horizon perturbations in the
slow-roll limit. Hence we will only consider super-horizon scales, and while
we will leave the scale factor unspecified in this subsection, it is worth
keeping in mind that this analysis may not be valid in spacetimes
that are not quasi-de Sitter. Using this solution for $\xi$, we can write the $0-i$
Einstein equation to $\mathcal{O}(m/M_{\mathrm{Pl}})$ as 
\begin{equation}
\Phi ^{\prime }+{\mathcal{H}}\Phi =4\pi G\left( \bar{\phi}^{\prime }+ma\bar{V
}_{\theta \phi }\right) \delta \phi .  \label{eq:0iom1}
\end{equation}
It is an interesting result that we can write the $0-i$ Einstein equation in geometrical terms as 
\begin{equation}
\Phi ^{\prime }+{\mathcal{H}}\Phi =A\delta \phi /\bar{\phi}^{\prime }
\label{eq:0iom1Aform}
\end{equation}
to both zeroth and first order in $\mathcal{O}(m/M_{\mathrm{Pl}})$. This
does not hold, however, to higher orders, and might not hold away from
quasi-de Sitter spacetime or on sub-horizon scales.

Next we solve the metric perturbation $\Phi $ to $\mathcal{O}(m/M_{\mathrm{Pl
}})$. Our master equation is the sum of the $0-0$ and $i-i$ Einstein
equations, dropping a $k^{2}\Phi $ term which is negligible on super-horizon
scales, 
\begin{align}
-8\pi Ga^{2}\bar{V}_{\phi }\delta \phi ={}& \Phi ^{\prime \prime }+6{
\mathcal{H}}\Phi ^{\prime }+2\left( 3{\mathcal{H}}^{2}-A\right) \Phi  \notag
\\
& +4\pi Gma\bar{V}_{\theta \phi }\left( \bar{\phi}^{\prime }\Phi -6{\mathcal{
H}}\delta \phi -\delta \phi ^{\prime }\right)  \notag \\
& -4\pi Gma\bar{V}_{\theta \phi \phi }\bar{\phi}^{\prime }\delta \phi + \ldots  \label{eq:00+ii}
\end{align}
where we have dropped terms at $\mathcal{O}(m/\Mp)^2$ and higher.

We want to remove the $\delta \phi $ terms from \cref{eq:00+ii} to write it
purely as an evolution equation for $\Phi $. To do this, we start with the
background relation (using \cref{eq:friedmann,eq:friedmann2}, assuming $\phi 
$ is gravitationally dominant) 
\begin{equation}
A={\mathcal{H}}^{2}-{\mathcal{H}}^{\prime }=4\pi G\left( \bar{\phi}^{\prime
2}+ma\bar{V}_{\theta \phi }\bar{\phi}^{\prime }\right) .
\end{equation}
Taking the conformal time derivative, we find (dropping $\mathcal{O}(m/M_{
\mathrm{Pl}})^{2}$ terms, as we do throughout) that 
\begin{equation}
\frac{A^{\prime }}{A}=\left( 2-\frac{ma\bar{V}_{\theta \phi }}{\bar{\phi}
^{\prime }}\right) \frac{\bar{\phi}^{\prime \prime }}{\bar{\phi}^{\prime }}+
\frac{ma\bar{V}_{\theta \phi }{\mathcal{H}}}{\bar{\phi}^{\prime }}+ma\bar{V}
_{\theta \phi \phi }.
\end{equation}
Using the background Klein-Gordon equation, we obtain 
\begin{equation}
-2a^{2}\bar{V}_{\phi }=\left( \frac{A^{\prime }}{A}+4{\mathcal{H}}\right) 
\bar{\phi}^{\prime }+ma\bar{V}_{\theta \phi }\left( \frac{1}{2}\frac{
A^{\prime }}{A}-{\mathcal{H}}\right) -ma\bar{V}_{\theta \phi \phi }\bar{\phi}
^{\prime }.  \label{eq:AVeq}
\end{equation}
In deriving the previous two expressions we have made use of the assumption that $ma\bar V_{\theta\phi}/\bar\phi' \sim \varepsilon^{-1/2}(m/\Mp)(\bar V_{\theta\phi}/H)$ is small compared to unity.

We will use \cref{eq:AVeq,{eq:0iom1}} to remove $\bar{V}_{\phi }$ and the $
\delta \phi $ terms from \cref{eq:00+ii}. We can also take the conformal
time derivative of \cref{{eq:0iom1}} to find (using \cref{eq:AVeq,{eq:0iom1}}
) an expression for $\delta \phi ^{\prime }$: 
\begin{equation}
4\pi G\bar{\phi}^{\prime }\delta \phi ^{\prime }=\Phi ^{\prime \prime
}+\left( {\mathcal{H}}-\frac{1}{2}\frac{A^{\prime }}{A}\right) \Phi ^{\prime
}+\left( {\mathcal{H}}^{2}-\frac{1}{2}\frac{A^{\prime }}{A}{\mathcal{H}}
-A\right) \Phi ,
\end{equation}
where we have dropped the $\mathcal{O}(m/M_{\mathrm{Pl}})$ term as $\delta
\phi ^{\prime }$ only appears in \cref{eq:00+ii} at that order.

Using these relations, as well as the definition of $A$, the sum of the $0-0$
and $i-i$ perturbed Einstein equations (\ref{eq:00+ii}) becomes 
\begin{align}
&\Phi^{\prime \prime }+ \left(2{\mathcal{H}} - \frac{A^{\prime }}{A}
\right)\Phi^{\prime }+ \left(2{\mathcal{H}}^2 -2A - \frac{A^{\prime }}{A}{
\mathcal{H}}\right)\Phi  \notag \\
={}&\frac{ma\bar V_{\theta\phi}}{\bar\phi^{\prime }}\left[\Phi^{\prime
\prime }+\left(2{\mathcal{H}}-\frac{A^{\prime }}{A}\right)\Phi^{\prime
}+\left(2{\mathcal{H}}^2-2A-\frac{A^{\prime }}{A}{\mathcal{H}}\right)\Phi
\right].
\end{align}
Simplifying, we find the evolution equation for $\Phi$ to $\mathcal{O}(m/M_
\mathrm{Pl})$, 
\begin{equation}
\boxed{\Phi'' + \left(2\sh - \frac{A'}{A}\right)\Phi' + \left(2\sh^2 -2A -
\frac{A'}{A}\sh\right)\Phi = 0. \label{eq:phievol}}
\end{equation}

This is a surprising result. This is exactly the equation obeyed by $\Phi $
in single-field slow-roll inflation in the absence of a coupling to any
other fields \cite{Mukhanov:1990me}. Coupling to new fields generically
introduces source terms to this equation, signalling the introduction of
isocurvature modes. We have shown that (to first order in $m/M_{\mathrm{Pl}
}$) the scalar-aether coupling does not produce any isocurvature modes on
super-horizon scales during slow-roll inflation.

What would happen if we included higher-order terms? The pure \ae -theory
terms do not change \cref{eq:phievol} \cite{Lim:2004js}. This is
understandable because the aether tracks the background energy density,
precluding the production of isocurvature modes. However, we have introduced
new coupling terms in the Einstein equations at $\mathcal{O}(m/M_{\mathrm{Pl}
})^{2}$, and higher, which could potentially produce isocurvature modes. It is currently unclear whether the unusual cancellations that led to the result (\ref{eq:phievol}) will hold at these orders.

The solution to \cref{eq:phievol} is well-known \cite{Mukhanov:1990me}, 
\begin{equation}
\Phi = C\left(1 - \frac{{\mathcal{H}}}{a^2}\int a^2d\tau\right),
\label{eq:phisoln}
\end{equation}
where $C$ is a constant. The remarkable fact that the $0-i$ Einstein
equation can be written in the form (\ref{eq:0iom1Aform}) to either zeroth
or first order in $m/M_\mathrm{Pl}$ means that to first order, the
relationship between $\Phi$ and $\delta\phi$ is the same as in the case
without the aether. Using \cref{eq:phisoln} to find $a^{-1}(a\Phi)^{\prime }$
and plugging that into \cref{eq:0iom1Aform}, we can determine the constant $
C $, 
\begin{equation}
C = \frac{aH}{\bar\phi^{\prime }}\delta\phi.
\end{equation}

The amplitude of $\delta \phi $ is determined by quantizing it in a
(quasi-)de Sitter background on sub-horizon scales, $k\gg aH$, and imposing a
Bunch-Davies vacuum state. $\delta \phi $ is coupled to the spin-0 aether
perturbations, as discussed in \cref{sec:flatspace}, and its dispersion
relation is modified by $\mu \equiv V_{\theta \phi }(0,0)$. However, the flat
spacetime stability condition constrains this to be less that the flat
spacetime mass of the scalar, $M_0 \equiv V_{\phi \phi }^{1/2}(0,0)$, up to an $
\mathcal{O}(1)$ factor. Therefore, if the initial conditions are set at
scales $k\gg M_0$ (which follows from $k\gg aH$ since $M_0 \ll H$), then $k\gg
\mu $ as well, and the scalar at these scales behaves as it does in the case
with no aether. We see that the scalar and metric perturbations, $\delta
\phi $ and $\Phi $, are the exact same as in general relativity to $\mathcal{
O}(m/M_{\mathrm{Pl}})$.

\subsection{The Large Coupling Limit: CMB Observables}

Let us finally connect these calculations to observations. As mentioned
at the beginning of this section, the two key inflationary observables currently accessible to CMB
experiments are the spectral index of the primordial power spectrum, $n_s$,
and the tensor-to-scalar ratio, $r$.

We have seen that, surprisingly, neither of these will be affected by the
aether-scalar coupling at $\mathcal{O}(m/M_\mathrm{Pl})$. Any new effects
must therefore enter at earliest at $\mathcal{O}(m/M_\mathrm{Pl})^2$. To
discuss these effects, we split $\Phi$ into zeroth-, first-, and
second-order pieces, 
\begin{equation}
\Phi = \Phi_\mathrm{GR} + \left(\frac{m}{M_\mathrm{Pl}}\right)^2\Phi_2 +
\ldots .  \label{eq:phiexpansion}
\end{equation}
Using this expansion, the power spectrum of $\Phi$ is 
\begin{equation}
P_\Phi = \langle\Phi^2\rangle = \langle\Phi_\mathrm{GR}^2\rangle + 2\left(
\frac{m}{M_\mathrm{Pl}}\right)^2\langle\Phi_\mathrm{GR}\Phi_2\rangle + \ldots.
\label{eq:phipowerspec}
\end{equation}

The deviation from scale-invariance, $n_s$, is defined by 
\begin{equation}
n_s - 1 = \frac{d \ln \Delta_\Phi^2}{d\ln k},
\end{equation}
where the dimensionless power spectrum is 
\begin{equation}
\Delta_\Phi^2 = \frac{k^3}{2\pi} P_\Phi.
\end{equation}
In GR, the deviation from scale-invariance is $-2\varepsilon-\eta$. Using
the results 
\begin{align}
\frac{d\ln\Phi_\mathrm{GR}^2}{d\ln k} &= -3-2\varepsilon-\eta, \\
\frac{d\ln\Phi_2^2}{d\ln k} &= -3 + \left(n_s-1\right)_2,
\end{align}
where $\left(n_s-1\right)_2$ is the spectral index of $\Phi_2$, and assuming
that $\Phi_2$ is not too much larger than $\Phi_\mathrm{GR}$, the spectral
index to second order in $m/M_\mathrm{Pl}$ is given by 
\begin{equation}
n_s - 1 = -2\varepsilon - \eta + \left(\frac{m}{M_\mathrm{Pl}}\right)^2\frac{
\Phi_2}{\Phi_0}\left[2\varepsilon + \eta + \left(n_s-1\right)_2\right] +
\ldots.
\end{equation}

Finally, we consider the tensor-to-scalar ratio, $r$, defined by 
\begin{equation}
r=\frac{\Delta _{t}^{2}}{\Delta _{\Phi }^{2}},
\end{equation}
where $\Delta _{t}^{2}$ is the dimensionless power spectrum of the spin-2
perturbations, $H_{Tk}^{(\pm 2)}$. Pure \ae -theory effects contribute a
constant rescaling to the tensor spectrum which only becomes important at $
\mathcal{O}(m/M_{\mathrm{Pl}})^{2}$ \cite{Lim:2004js}. Recall that the
coupling between the aether and $\phi $, however, has no effect on the
tensor perturbations as none of the coupling terms contain spin-2 pieces, so
the tensor spectrum $\Delta _{t}^{2}$ is unchanged apart from the
aforementioned (small) rescaling. Therefore, $r$ is modified by a factor 
\begin{equation}
\frac{r}{r_{\mathrm{GR}}}=\frac{\Delta _{\Phi _{0}}^{2}}{\Delta _{\Phi }^{2}}
,
\end{equation}
where $r_{\mathrm{GR}}$ is the tensor-to-scalar ratio in the absence of the
aether-inflaton coupling. Using the expansion (\ref{eq:phiexpansion}), we
find that the corrections to $r$ are small, 
\begin{equation}
\frac{r}{r_{\mathrm{GR}}}=1-2\left( \frac{m}{M_{\mathrm{Pl}}}\right) ^{2}
\frac{\Phi _{2}}{\Phi _{0}}+\ldots .
\end{equation}

What size are the corrections to $n_{s}-1$ and $r$? As
discussed in \cref{sec:lambdavalues}, $m/M_{\mathrm{Pl}}$ is no larger than $
\sim \mathcal{O}(10^{-7})$, barring any special cancellations among the $
c_{i}$. We constructed the expansion of $\Phi $ so that $\Phi _{2}$ is at
least not too much larger than $O(\Phi _{0})$. We assume that there are no
effects such as instabilities at $\mathcal{O}(m/M_{\mathrm{Pl}})$ which
would cause this construction to fail (the one instability that we have found
in the spin-0 modes, discussed in \cref{sec:spin0instability}, has been
assumed to vanish, by making the coupling either very small or of the opposite sign to $\dot{\bar\phi}$). The Planck sensitivity to $r$ is about $10^{-1}$, and
about $10^{-2}$ to $n_{s}-1$ \cite{Ade:2013zuv,Ade:2013uln}.

We see that the first corrections to $\Phi$ enter at $\mathcal{O}(m/M_
\mathrm{Pl})^2$. This is constrained by other experiments to be a tiny
number, placing any coupling between $\phi$ and $\theta$ which is not
already ruled out far outside the current and near future window of CMB
observability.

\section{Case Study: Quadratic Potential}

\label{sec:dj}

\subsection{Slow-Roll Inflation: An Example}

\label{sec:djsr}

The arguments so far have been made for a general potential $V(\theta,\phi)$ with
only minimal assumptions. In order to be more quantitative, we will now look
more closely at a particular form of the potential for which the
inflationary dynamics are known and relatively simple.

The Donnelly-Jacobson potential \cite{Donnelly:2010cr} contains all terms
relevant to the dynamics at quadratic order in the fields and is given by 
\begin{equation}
V(\theta ,\phi )=\frac{1}{2}M^{2}\phi ^{2}+\mu \theta \phi .  \label{eq:DJ}
\end{equation}
A term proportional to $\theta $ contributes a total derivative to the
action and hence is non-dynamical (note that the potential enters the
Friedmann equation through $V-\theta V_{\theta }$, not $V$ itself), while a
term proportional to $\theta ^{2}$ can be absorbed into $c_{2}$ and would
only renormalize $G_{c}$. We take $\mu >0$ as the theory is invariant under
the combined symmetry $\mu \rightarrow -\mu $ and $\phi \rightarrow -\phi $.
Any dynamics with $\mu <0$ can be obtained by flipping the sign of $\phi $.

This is simple $m^2\phi ^{2}$ chaotic inflation with an extra force that pushes 
$\phi $ towards negative values \cite{Donnelly:2010cr}. In the case where the scalar field has no mass term, $\phi $ possesses exact shift symmetry, $\phi \rightarrow \phi +
\mathrm{const.},$ and this theory is essentially $\Theta \mathrm{CDM}$, a
dark energy theory in which $\mu $ is related to the dark energy scale and,
importantly, is protected from radiative corrections by the existence of a
discrete symmetry \cite{Blas:2011en,Audren:2013dwa}. Interestingly, in the
special case where the aether is hypersurface-orthogonal, this theory also
admits a candidate UV completion in the consistent non-projectable extension 
\cite{Blas:2009qj,Blas:2009ck,Blas:2010hb} of Ho\v{r}ava-Lifschitz gravity 
\cite{Horava:2009uw}. In that case, however, the spin-1 modes we have
discussed vanish. This is because the aether can be written as the
(normalized) gradient of a scalar field corresponding to a global time
coordinate, so it possesses no spin-1 modes. A similar coupling was also
considered in \cite{Libanov:2007mq}.

The equations of motion, in conformal time, are 
\begin{align}
{\mathcal{H}}^{2}& =\frac{4\pi G_{c}}{3}a^{2}\left( M^{2}\phi ^{2}+\phi
^{\prime 2}a^{-2}\right) ,  \label{eq:djfriedmann} \\
{\mathcal{H}}^{\prime }& =\frac{4\pi G_{c}}{3}a^{2}\left( M^{2}\phi
^{2}-2\phi ^{\prime 2}a^{-2}-3\frac{m\mu }{a}\phi ^{\prime }\right) , \\
0& =\phi ^{\prime \prime }+2{\mathcal{H}}\phi ^{\prime 2}M^{2}\phi +3{
\mathcal{H}}m\mu a.  \label{eq:kg}
\end{align}
Normally, we can obtain a slow-roll inflationary solution to leading order
by neglecting $\dot{\phi}^{2}=a^{-2}\bar{\phi}^{\prime 2}$ in the Friedmann
equation \cref{eq:djfriedmann} and $\ddot{\bar{\phi}}$\footnote{We cannot just drop $\bar{\phi}^{\prime \prime }$ as it contains a term like 
$H\dot{\phi}$. It is easiest to drop the second derivative piece from the
cosmic time scalar evolution equation and \textit{then} move to conformal
time.} in the scalar evolution equation \cref{eq:kg}. The same applies in
this theory; we now briefly justify this.

A slow-roll inflationary phase requires $H$ to be changing slowly, and for
inflation to be successful it needs to last at least 50-60 $e$-folds.
This is guaranteed by making sure the slow-roll parameters 
\begin{alignat}{2}
\varepsilon & =-\frac{\dot{H}}{H^{2}} & & =1-\frac{{\mathcal{H}}^{\prime }}{{
\mathcal{H}}^{2}}, \\
\eta & =\phantom{-}\frac{\dot{\varepsilon}}{H\varepsilon } & & =\frac{
\varepsilon ^{\prime }}{{\mathcal{H}}\varepsilon },
\end{alignat}
are both very small compared to unity. For convenience we will work in
cosmic time ($t=\int ad\tau $) here. The slow-roll parameters are 
\begin{align}
\varepsilon & =\frac{4\pi G_{c}}{H^{2}}\left( \dot{\phi}^{2}+m\mu \dot{\phi}
\right) , \\
\eta & =2\left[ \varepsilon +\frac{\ddot{\phi}}{H\dot{\phi}}\left( \frac{2
\dot{\phi}+m\mu }{2\dot{\phi}+2m\mu }\right) \right] .
\end{align}
Defining 
\begin{align}
\delta & \equiv \frac{4\pi G_{c}\dot{\phi}^{2}}{3H^{2}}, \\
\lambda & \equiv \frac{\ddot{\phi}}{3H\dot{\phi}}, \\
\gamma & \equiv \frac{M}{H}\frac{\mu }{\mu _{c}},
\end{align}
where 
\begin{equation}
\mu _{c}\equiv \frac{1}{\sqrt{12\pi G_{c}}}\frac{M}{m},  \label{eq:mucdef}
\end{equation}
we can calculate the slow-roll parameters, 
\begin{align}
\varepsilon & =3\delta +\gamma \delta ^{1/2}, \\
\eta & =-3\lambda \left( \frac{6\delta ^{\frac{1}{2}}+\gamma }{6\delta ^{
\frac{1}{2}}+2\gamma }\right) .
\end{align}

The usual slow-roll conditions $\dot{\phi}^{2}\ll H^{2}$ and $\ddot{\phi}\ll
3H\dot{\phi}$ are equivalent to $\delta \ll 1$ and $\lambda \ll 1$,
respectively. We generally expect $M<H$ in order for the inflaton to produce
perturbations. As we will see below, the stability considerations discussed
in \cref{sec:flatspace} require $\mu <\mu _{c}$. When combined, these
conditions imply $\gamma <1$. So, under these reasonable assumptions on $M$
and $\mu $, in order to ensure $\varepsilon \ll 1$ and $\eta \ll 1,$ we
simply need $\dot{\phi}^{2}\ll H^{2}$ and $\ddot{\phi}\ll 3H\dot{\phi}$ as
usual. Note, however, that the usual identifications of $\varepsilon $ and $
\eta $ in terms of the potential will be changed if the scalar-aether
coupling is large enough for $\gamma $ to be comparable to $\delta ^{1/2}$.

In the slow-roll limit, the Friedmann and Klein-Gordon equations are,
respectively, 
\begin{align}
\mathcal{H} &  \simeq\sqrt{\frac{4\pi G_{c}}{3}}Ma|\bar{\phi}
|,\label{eq:shslowroll}\\
\phi^{\prime} &  \simeq-\sqrt{\frac{1}{12\pi G_{c}}}Ma\left(
\operatorname{sgn}(\phi)+\frac{\mu}{\mu_{c}}\right)
.\label{eq:phiprimeslowroll}
\end{align}
Notice the appearance of $\mu_{c}$ defined above. During slow-roll, it is
related to the inflationary dynamics by 
\begin{equation}
\mu_{c}=\frac{M^{2}|\phi|}{\theta}.
\end{equation}

The value of $\mu /\mu _{c}$ is physically significant because it determines
the stability of the slow-roll solution. The number of $e$-folds that
inflation lasts tends to infinity as $\mu \rightarrow \mu _{c}$, which
corresponds to exact de Sitter expansion; for $\mu >\mu _{c}$ the slow-roll
solution is unstable and grows without bound \cite{Donnelly:2010cr}. Hence,
we will always consider inflationary solutions with $\mu <\mu _{c}$.

There is an additional constraint on $\mu/\mu_{c}$ from the spin-0 stability
constraint (\ref{eq:spin0constraint}). Substituting the definition of $\mu
_{c}$ into this gives the constraint 
\begin{equation}
\frac{\mu^{2}}{\mu_{c}^{2}}\leq24\pi G_{c}m^{2}c_{123}=\frac{24\pi
Gm^{2}c_{123}}{1+8\pi G\alpha}.  \label{eq:muconstraint}
\end{equation}
The same constraint was derived along similar lines in \cite{Donnelly:2010cr}.\footnote{As mentioned in \cref{sec:flatspace}, our action and potential differ from
those in \cite{Donnelly:2010cr}, because we give the aether units of mass
while their aether is dimensionless. Taking the different definitions of $
c_{i}$, $m$, and $\mu$ into account, our constraint agrees with theirs.}
Since $c_{123}\leq1$ and $\alpha\geq0$ (see \cref{sec:flatspace}, as well as 
\cite{Lim:2004js,Carroll:2004ai}), this is more restrictive than simply $
\mu<\mu_{c}$, unless $m$ is comparable to, or greater than, the Planck scale
-- a possibility that seems to be ruled out by experiments, as discussed in 
\cref{sec:lambdavalues}. Since experiments suggest $m/\Mp \lesssim 10^{-7}$, $\mu/\mu_c$ must be so small that inflationary dynamics would be effectively unchanged by the coupling, unless cancellations among the $c_i$ conspire to weaken the bounds on $m$.

\subsection{The Instability Explored}

Specializing to the Donnelly-Jacobson potential, and using the slow-roll
equations (\cref{eq:shslowroll,eq:phiprimeslowroll}) we can write the spin-1
equation of motion (\ref{eq:spin1xislowroll}) to first order in the
slow-roll parameters as 
\begin{equation}
\xi^{\prime\prime}+c_{s}^{{(\pm1)}2}k^{2}\xi-\frac{\Lambda}{\tau^{2}}\xi=0,
\label{eq:djxieq}
\end{equation}
with $\Lambda$ given by 
\begin{align}
\Lambda & \equiv -\frac{1}{2}\frac{\mu\mu_{c}}{c_{1}H^{2}}\left(
\operatorname{sgn}(\phi)+\frac{\mu}{\mu_{c}}\right)  +\mathcal{O}
(\varepsilon),\nonumber\\
&  =-\frac{M^{2}}{H^{2}}\left(  c_{s}^{(0)2}\sigma+\operatorname{sgn}(\mu
\phi)\frac{c_{s}^{(0)}\sqrt{\sigma}}{\sqrt{3c_{1}}}\sqrt{\frac{M_{\mathrm{Pl}
}^{2}}{m^{2}}+c_{13}+3c_{2}}\right)  +\mathcal{O}(\varepsilon
).\label{eq:lambdadefdj}
\end{align}
Here, as in \cref{sec:lambdavalues}, we have defined the normalized coupling 
$\sigma$ by 
\begin{equation}
\mu^{2}=2c_{123}M^{2}\sigma=24\pi G_{c}m^{2}c_{123}\sigma\mu_{c}^{2},
\label{eq:djsigdef}
\end{equation}
so that flat spacetime stability of the spin-0 modes implies $\sigma\leq1$.

As with the general case, the solution (\ref{eq:vksol}) to \cref{eq:djxieq}
is written in terms of the first Hankel function of order $\nu$, where 
\begin{equation}
\nu^{2} \equiv\frac{1}{4} + \Lambda
\end{equation}

Repeating the analysis of \cref{sec:instability}, we pick a single mode
which leaves the sound horizon at some conformal time $\tau _{i}$, which we
could take to be the start of inflation. We pick a mode which crosses the
horizon early because $V_{k}(\tau )$ is largest at small $k$ (with $\tau $
held fixed), so this is one of the larger super-horizon modes available. We
want to calculate the contribution of this mode to the spacetime components
of the stress-energy tensor. If it exceeds the background energy density,
then this would indicate a violation of expansion isotropy and signal an
instability in the background solution, as we found in \cref{sec:djsr}.

Using the slow-roll scalar equation, and our expression (\ref
{eq:phiprimeslowroll}) for $\bar{\phi}^{\prime }$, we find 
\begin{align}
\bar{V}_{\phi }\bar{V}_{\theta \phi }& =M\mu H\sqrt{\frac{3}{4\pi G_{c}}}
\left( \operatorname{sgn}(\phi )+\frac{\mu }{\mu _{c}}\right)  \notag \\
& =M^{2}H\left( 6mc_{123}\sigma +\operatorname{sgn}(\mu \phi )\sqrt{12c_{123}\sigma 
\left[ M_{\mathrm{Pl}}^{2}+(c_{13}+3c_{2})m^{2}\right] }\right) .
\end{align}
We can substitute this directly into \cref{eq:t0icomp} to find one of the
terms in the contribution that this mode makes to $T^{0}{}_{i}$, 
\begin{equation}
T^{0}{}_{i,k}/\bar{T}^{0}{}_{0}\supset \frac{c_{s}^{(0)}}{12\sqrt{3}\pi }
\frac{M}{H}\frac{M}{\sqrt{M_{\mathrm{Pl}}^{2}+\alpha }}\left( \operatorname{sgn}(\mu
\phi )\sqrt{\sigma }+\sqrt{3c_{123}}\sigma \frac{m}{\sqrt{M_{\mathrm{Pl}
}^{2}+\alpha }}\right) \Gamma (\nu )2^{\nu -\frac{1}{2}}\left( -\tau
_{i}\right) ^{\frac{3}{2}}e^{\left( \nu -\frac{3}{2}\right) N}\delta
^{3}{}_{i}.  \label{eq:djt0icomp}
\end{equation}

We can now get a more quantitative handle on the argument made in 
\cref{sec:instability}. Assuming $\nu >3/2$, the exponential in $(\nu -3/2)N$
is likely to overwhelm the other terms within the 50--60 or more $e$-folds
that will occur after $\tau _{i},$ which we take to be near the start of
inflation. While several terms in \cref{eq:djt0icomp} are likely to be
several orders of magnitude smaller than unity, including $M/H$, $m/M_{
\mathrm{Pl}}$,\footnote{A requirement for $\nu $ to be greater than $3/2$ in the first place.} and
possibly $M/M_{\mathrm{Pl}}$, it is unlikely that these could be so small as
to overwhelm the exponential terms and the gamma function. Hence, for $\nu
>3/2,$ we expect that the slow-roll background solution we found in 
\cref{sec:djsr} is unstable, rapidly dominated by perturbations in the
aether field generated by its coupling to the inflaton.

In \cref{sec:lambdavalues} we found that $\nu$ can surpass $3/2$, even by
several orders of magnitude, if the aether VEV, $m,$ is suitably small
compared to the Planck scale. Armed with a specific form for the potential,
we now briefly clarify that argument.

If $\nu >3/2$ then $\Lambda >2$, where $\Lambda $ is defined in 
\cref{eq:lambdadefdj}. It is not difficult to check that this is the same as
the $\Lambda $ we discussed for a general potential, \cref{eq:lambdadef},
which we wrote in various forms in \cref{sec:lambdavalues}. There, we found
that for $\Lambda $ to be positive we needed $\mu \dot{\phi}$ to be
positive. With the Donnelly-Jacobson potential, we have an expression for $
\dot{\phi}$, \cref{eq:phiprimeslowroll}. From that we see that $\mu \dot{\phi
}$ is only positive (assuming $\mu <\mu _{c}$) when $\mu \phi $ is negative.
We will take $\mu $ to be positive and then ask if $\phi $ can be negative
(the opposite case is trivial, as the theory has combined $\mu \rightarrow
-\mu $, $\phi \rightarrow -\phi $ symmetry). This is not at all uncommon,
and depends only on initial conditions. The dynamics for this inflationary
model are encapsulated in $(\phi ,\dot{\phi})$ phase portraits for a range
of $\mu /\mu _{c}$ in \cite{Donnelly:2010cr}. Per \cref{eq:djsigdef}, $\mu
/\mu _{c}$ is of order $(m/M_{\mathrm{Pl}})\sigma ^{1/2}$. Because
observations suggest $m\ll M_{\mathrm{Pl}}$ (see \cref{sec:lambdavalues}), $
\mu $ should be very small compared to $\mu _{c}$ even when $\sigma $
approaches unity. Hence, the phase portrait for $\mu =0$ in \cite
{Donnelly:2010cr} will be very close to the dynamics we are interested in.
In the exact $\mu =0$ case, there are as many inflating paths with $\phi <0$
as $\phi >0$, because when $\mu =0$, the equations for $\phi $ and $\dot{\phi
}$ have combined $\phi \rightarrow -\phi $ and $\dot{\phi}\rightarrow -\dot{
\phi}$ symmetry. The next phase portraits show a tendency, increasing with $
\mu $, for inflating paths to live in the $\phi >0$ half of the phase plane.
Since $\mu \ll \mu _{c}$, nearly half of all initial conditions leading to
viable inflation have $\mu \phi <0$.

Considering each piece in $\Lambda$ on an order-of-magnitude basis, and
taking $\operatorname{sgn}(\mu\phi)=-1$, we have 
\begin{equation}
\Lambda = -\underbrace{\frac{M^{2}}{H^{2}}}_{\ll1}\left( \underbrace {
c_{s}^{(0)2}\sigma\vphantom{\frac{c_s^{(0)}\sqrt{\sigma}}{\sqrt{3c_1}}}}
_{\lesssim1}-\underbrace{\frac{c_{s}^{(0)}\sqrt{\sigma}}{\sqrt{3c_{1}}}}_{
\mathcal{O}(1)}\sqrt{\frac{M_{\mathrm{Pl}}^{2}}{m^{2}}+c_{13}+3c_{2}}\right)
+\mathcal{O}(\varepsilon).
\end{equation}
Evidently, $\Lambda$ will be greater than $2$ if the smallness of $m$
compared to the Planck scale exceeds the (square of the) smallness of the
scalar mass, $M$, compared to the Hubble scale, 
\begin{equation}
\frac{M_{\mathrm{Pl}}}{m}\gtrsim\frac{2\sqrt{3c_{1}}}{c_{s}^{(0)}\sqrt{
\sigma }}\left( \frac{M}{H}\right) ^{-2},
\end{equation}
where we have assumed that $m/\Mp\ll1$. While $M/H$ should be small, there are no limits on how small $m/M_{\mathrm{
Pl}}$ should be before the collider scale, and moreover, as discussed in 
\cref{sec:lambdavalues}, there are already likely to be stringent
experimental constraints on $m/M_{\mathrm{Pl}}$ (although these tend to
depend on the $c_{i}$ not cancelling out in particular ways).

The tachyonic instability discussed here and in \cref{sec:spin1perts} is absent when $\mu$ and $\phi$ have the same sign. In this case, the coupling only serves to dampen aether perturbations. For the Donnelly-Jacobson potential, what remains is effectively just $m^2\phi^2$ inflation. If the signs of $\mu$ and $\phi$ are different, or if we were to demand that inflation be viable for all initial conditions, then the absence of this instability puts a very strong constraint on the magnitude of $\mu$,
\begin{equation}
\boxed{\frac{|\mu|}{H} \lesssim 2\sqrt{6}c_1\frac{m}{\Mp}\left(\frac{M}{H}\right)^{-2}.}
\end{equation}
From the background dynamics, we expect $(M/H)^2 \approx 3\varepsilon + \mathcal{O}(m/\Mp) \sim \mathcal{O}(10^{-2})$, while the absence of gravitational \v{C}erenkov radiation constrains $m/\Mp \lesssim \mathcal{O}(10^{-7})$, in the absence of certain cancellations among the $c_i$. Thus the constraint on $\mu$ is of the order
\begin{equation}
\frac{|\mu|}{H} \lesssim \mathcal{O}(10^{-5}).
\end{equation}
This should be compared to the previous strongest constraint on $\mu$, the flat spacetime stability constraint discussed in \cite{Donnelly:2010cr} and \cref{sec:flatspace},
\begin{equation}
\frac{|\mu|}{H} < \sqrt{2c_{123}} \sim \mathcal{O}(1).
\end{equation}

\section{Discussion}

\label{sec:discussion}

We have examined cosmological perturbations in a theory of single-field,
slow-roll inflation coupled to a vector field that spontaneously breaks
Lorentz invariance, looking both to explore the effects of such a coupling on inflationary cosmology and to place constraints on it. The particular model is Einstein-aether theory, a
theory of a fixed-norm timelike vector called the \textquotedblleft aether,"
coupled to a canonical scalar field by allowing its potential to depend on the
divergence of the aether, $\theta =\nabla _{\mu }u^{\mu }$. In a homogeneous
and isotropic cosmology, $\theta$ is related to the Hubble rate, $H=\theta
/3m$. This construction allows $H$ to play a role in cosmological dynamics
that it cannot in general relativity, where it is not a spacetime scalar. Moreover, it is a fairly general model of coupling between a fixed-norm vector and a scalar field. In particular, while many couplings can be written down which are not captured by a potential $V(\theta,\phi)$, all such terms have mass dimension 5 or higher and therefore would not be power-counting renormalizable.

Around a slow-roll inflationary background, this theory possesses a tachyonic instability. The instability is present if the norm of the aether, effectively the Lorentz symmetry breaking scale, is sufficiently small compared to the Planck mass, and the aether-scalar coupling is suitably large. In this region of parameter space, assuming a technical requirement on the initial conditions, scalar and vector perturbations both grow exponentially, destroying the inflationary background. Demanding the absence of this instability for general initial conditions places a constraint on the coupling which is significantly stronger than the existing constraints, which are based on stability of the perturbations around flat spacetime and viability of a slow-roll solution. Hence this constraint is by far the strongest on an aether-scalar coupling to date, with the assumption that the scalar drives a slow-roll inflationary period.

The root of the instability is the smallness of the aether VEV, $m$, compared to the Planck mass. The non-coupled terms in the aether Lagrangian each have two factors of $u^\mu$, so these aether terms will come with a factor of $(m/\Mp)^2$ in the Einstein equations. Terms involving two or more $\theta$ derivatives of the scalar field potential will also enter the Einstein equations with these factors or higher. However, terms associated with the coupling $V_{\theta\phi}$, which only has one aether derivative, will only have one power of $m/\Mp$ and so will generically be larger (depending on the size of $V_{\theta\phi}$) than the other aether-related terms. In the aether equation of motion, this coupling term will be a power of $\Mp/m$ larger than the other terms for the same reason. When the coupling is sufficiently large, it is exactly this term that drives the instability.

If the instability is absent, then observables in the CMB are unaffected by the coupling at the level of observability of current and near-future experiments; the corrections are smaller than $\mathcal{O}(10^{-15})$. This is due partly to the smallness of the aether norm relative to the Planck scale, but is exacerbated by the presence of unusual cancellations. Solving for the spin-0 perturbations order by order in the aether VEV, $m/\Mp$, no isocurvature modes are produced at first order. This is unexpected, as isocurvature modes are a generic feature of multi-field theories. Stronger yet, the perturbations are completely unchanged at first order in $m/\Mp$ from the case without any aether at all. This is largely a result of unexpected cancellations which hint at a deeper physical mechanism. An explanation of such a mechanism is left to future work.

Also left to future work is whether these unexpected conclusions hold to higher orders in $m/\Mp$. At $\mathcal{O}(m/\Mp)^2$, several new coupling terms enter the perturbed Einstein equations, \cref{eq:spin000,eq:spin00i,eq:spin0ij}, with a qualitatively different structure to the terms which appear at $\mathcal{O}(m/\Mp)$. The possibility therefore remains that the isocurvature modes that one would expect from the multiple interacting scalar degrees of freedom might re-emerge at this level. If they do, they would be severely suppressed relative to the adiabatic modes.

Beyond perhaps an extreme fine-tuning, there does not seem to be a subset of the parameter space in which observable vector perturbations are produced without destroying inflation. Even if such modes could be produced, they do not freeze out on super-horizon scales and are sensitive to the uncertain physics, such as reheating, between the end of inflation and the beginning of radiation domination. Therefore any observational predictions for vector modes would be strongly model-dependent. Nonetheless, it should be stressed that the line between copious vector production (that quickly overcomes the background) and exponentially decaying vector production is so thin, as it depends on unrelated free parameters, that there is no reason to expect this theory would realize it.

While we made these arguments for a general potential, we also looked at a
specific, simple worked example, the potential of Donnelly and Jacobson \cite{Donnelly:2010cr}. This potential includes all dynamical terms at quadratic order, and amounts to $m^2\phi^2$ chaotic inflation with a coupling to the aether that
provides a driving force. It contains many of the terms allowed for the aether and scalar up to dimension 4.\footnote{One could also add a tadpole term proportional to $\phi$ and a term proportional to $\phi^2\theta$. The latter would effectively promote the coupling $\mu$ to $\mu + \mathrm{const.}\times\phi$, so during slow-roll inflation the effective $\mu$ would still be roughly constant.} The constraint this places on the coupling $\mu \equiv V_{\theta\phi}$,
\begin{equation}
\frac{|\mu|}{H} \lesssim 2\sqrt{6}c_1\frac{m}{\Mp}\left(\frac{M}{H}\right)^{-2} \lesssim \mathcal{O}(10^{-5}),
\end{equation}
is stronger by several orders of magnitude than the the next best constraint \cite{Donnelly:2010cr},
\begin{equation}
\frac{|\mu|}{H} < \sqrt{2c_{123}} \sim \mathcal{O}(1).
\end{equation}
It is worth emphasizing again the two conditions for our constraint to hold. First, the scalar must drive a period of slow-roll inflation. Second, the instability can be avoided if $\mu$ and $\phi$ have the same sign. Consequently, the new constraint applies only if we demand that inflation be stable for all initial conditions. Assuming such a coupling exists, this constraint could be seen as a lower bound on $m$, to be contrasted to the many upper bounds on $m$ in the literature.

\begin{acknowledgments}
ARS is grateful to Neil Barnaby, Daniel Baumann, Anne Davis, Alec Graham,
Eugene Lim, Jeremy Sakstein, and Sergey Sibiryakov for helpful discussions. We also thank the referee for several useful comments and suggestions. ARS is supported by the David Gledhill Research Studentship, Sidney Sussex
College, University of Cambridge; and by the Isaac Newton Fund and
Studentships, University of Cambridge. JDB is supported by the STFC.
\end{acknowledgments}

\appendix

\section{Real Space Cosmological Perturbation Equations}

\label{app:realspace}

In this appendix, we present the real space equations of motion for the
cosmological perturbations.

We have for the $\nu=0$ component of the aether field equation (\ref
{eq:aethereom})
\begin{align}
&-6(c_{13}+2c_2)\sh^2\Phi+6c_2\left(\frac{a''}{a}\right)\Phi \nonumber \\
&+\sh\left[(2c_1+c_2)V^i{}_{,i}+c_3(V^i{}_{,i}+ B^i{}_{,i})+3c_2\Phi'+3(2c_{13}+c_2)\Psi'\right] \nonumber \\
&-c_3(\Phi^{,i}{}_i-B{}^i{}_{,i}'+V{}^i{}_{,i}')-c_2(V^i{}_{,i}'+3\Psi'')+a^2\delta \lambda_\textrm{\ae} \nonumber \\ 
&+ \frac{1}{2}\bar V_{\theta\theta}\left[6\left(\frac{a''}{a}-2\mathcal{H}^2\right)\Phi + 3\mathcal{H}(\Phi^\prime+\Psi^\prime)-3\Psi''+\mathcal{H}V^i{}_{,i} - V^i{}_{,i}^\prime \right] \nonumber \\
& - \frac{3}{2}\frac{m}{a}\bar V_{\theta\theta\theta}\left(\frac{a''}{a}-2\mathcal{H}^2\right)\left(3\Psi^\prime-3\mathcal{H}\Phi+V^i{}_{,i}\right) \nonumber \\
& + \frac{1}{2}\frac{a}{m}\left[\bar V_{\theta\phi}(\bar\phi'\Phi -\delta\phi') - \bar V_{\theta\phi\phi}\bar\phi'\delta\phi\right] \nonumber \\
& - \frac{1}{2}\bar V_{\theta\theta\phi}\left[\bar\phi'(3\Psi' - 3\mathcal{H}\Phi + V^i{}_{,i}) + 3\left(\frac{a''}{a}-2\mathcal{H}^2\right)\delta\phi \right] = 0
\end{align}
and the $\nu=i$ component is
\begin{align}
&-\left[2\frac{\alpha}{m^2}\sh^2-\frac{\alpha}{m^2}\left(\frac{a''}{a}\right)+c_1\left(\frac{a''}{a}\right)\right](B_i-V_i) \nonumber \\
&+\sh\left[(c_1+\frac{\alpha}{m^2})\Phi_{,i}+2c_1(V_i'-B_i')\right] \nonumber \\
&+\frac{1}{2}(-c_3+c_1)B_{[i,j]}{}^j-c_1V_{i,j}{}^j-c_{23}V^j{}_{,ij} \nonumber \\
&-c_{13} h'{}^j{}_{i,j}+c_1\Phi'_{,i}-\frac{\alpha}{m^2} \Psi'_{,i}-c_1(B_i''-V_i'') \nonumber \\
& +\frac{1}{2}\bar V_{\theta\theta}\left[3\left(\frac{a''}{a}-2\mathcal{H}^2\right)(B_i-V_i)-3\Psi^\prime_{,i}+3\mathcal{H}\Phi_{,i}-V^j{}_{,ij}\right] \nonumber \\
& + \frac{1}{2}\frac{a}{m}\bar V_{\theta\phi}\left(\bar\phi'(B_i-V_i)-\delta\phi_{,i}\right)=0
\end{align}

The combined aether-scalar stress energy tensor (\ref{eq:fullsetensor}) has
perturbations
\begin{align}
\delta T^0{}_0 ={}& 2\frac{m^2}{a^2}\Big\{-3(c_{13}+3c_2)\sh^2\Phi+c_1a^{-1}\left[a(B^i-V^i)_{,i}\right]' \nonumber \\
& +(c_{13}+3c_2)\sh(V^i{}_{,i}+3\Psi')+c_1\Phi^{,i}{}_i\Big\} \nonumber \\
& + \frac{1}{a^2}\left(\bar\phi'^2\Phi-\bar\phi'\delta\phi'-a^2\bar V_\phi\delta\phi\right) \nonumber \\
&+ \frac{3m^2}{a^2}\mathcal{H}\bar V_{\theta\theta}\left(3\Psi^\prime-3\mathcal{H}\Phi+V^i{}_{,i}\right) + \frac{3m}{a}\mathcal{H}\bar V_{\theta\phi}\delta\phi \\
\delta T^0{}_i ={}& 2\frac{m^2}{a^2}\left\{\left[-2(c_{13}+3c_2)\sh^2+(3c_2+c_3)\left(\frac{a''}{a}\right)\right](V_i-B_i) \right. \nonumber \\
&-c_1a^{-2}\left[a^2(V_i'-B_i')\right]'-c_1a^{-1}(a\Phi_{,i})' \nonumber \\
& \left. +\frac{1}{2}(-c_1+c_3)\left[(B_i-V_i)^{,j}{}_j-(B^j-V^j)_{,ij}\right]\right\} \nonumber \\
& - \frac{1}{a^2}\bar\phi'\delta\phi_{,i} + \frac{3m^2}{a^2}\bar V_{\theta\theta}\left(\frac{a''}{a}-2\mathcal{H}^2\right)(V_i-B_i) + \frac{m}{a}\bar V_{\theta\phi}\bar\phi'(V_i-B_i) \\
\delta T^i{}_j ={}& 2\frac{m^2}{a^2}\left\{(c_{13}+3c_2)\left[\sh^2-2\left(\frac{a''}{a}\right)\right]\Phi\delta^i{}_j-(c_{13}+3c_2)\sh\Phi'\delta^i{}_j \right. \nonumber \\
& +\left.a^{-2}\bigg[a^2(c_2V^k{}_{,k}\delta^i{}_j+(c_{13}+3c_2)\Psi'\delta^i{}_j +\frac{1}{2}c_{13}(V^i{}_{,j}+V_j{}^{,i}+2h^i{}_j')\bigg]'\right\} \nonumber \\
& + \left\{- \frac{1}{a^2}\left(\bar\phi'^2\Phi-\bar\phi'\delta\phi'+a^2\bar V_\phi\delta\phi\right)\right. \nonumber \\
& + \left.\frac{m^2}{a^2}\bar V_{\theta\theta}\left[3\left(\mathcal{H}^2-2\frac{a''}{a}\right)\Phi-3\mathcal{H}\Phi^\prime+a^{-2}\left(a^2\left(3\Psi^\prime+V^k{}_{,k}\right)\right)^\prime\right]\right. \nonumber\\
& + \left.\frac{3m^3}{a^3}\bar V_{\theta\theta\theta}\left(\frac{a''}{a}-2\mathcal{H}^2\right)\left(3\Psi^\prime-3\mathcal{H}\Phi+V^{k}{}_{,k}\right)\right. \nonumber \\
& + \left.\frac{m}{a}\left[\bar V_{\theta\phi}(3\mathcal{H}\delta\phi+\delta\phi'-\bar\phi'\Phi) + \bar V_{\theta\phi\phi}\bar\phi'\delta\phi\right]\right. \nonumber \\
& + \left. \frac{m^2}{a^2}\bar V_{\theta\theta\phi}\left(3\left(\frac{a''}{a}-2\mathcal{H}^2\right)\delta\phi+\bar\phi'(3\Psi'-3\mathcal{H}\Phi+V^k{}_{,k})\right)\right\}\delta^i{}_j.
\end{align}

We can do a consistency check by choosing $V(\theta,\phi) = \frac{1}{2}
\beta\theta^2 + V(\phi)$. This corresponds to pure \ae -theory, with $c_2$
rescaled to $c_2 + \beta$, and a scalar field coupled only to gravity. The
cosmological perturbations in that model are presented in \cite{Lim:2004js}.
Our equations agree with the literature in this limit, as we would expect.

\bibliography{bibliography}

\end{document}